\documentclass[a4paper,10pt,notitlepage]{article}
\pdfoutput=1
\usepackage{ulem}
\usepackage{datetime}
\usepackage{hyperref}
\usepackage{amsfonts,amssymb,amsmath,amsthm,colonequals}
\usepackage[mathscr]{euscript}
\usepackage{bbm}
\usepackage[show]{ed}
\usepackage{makeidx}
\usepackage{graphicx}
\usepackage{url}
\usepackage{multirow}
\usepackage{youngtab}
\usepackage[arrow, matrix, curve]{xy}
\usepackage{simplewick}
\theoremstyle{definition}
\newcommand{\beqa}{\begin{eqnarray}}
\newcommand{\eeqa}{\end{eqnarray}}
\newcommand{\beq}{\begin{equation}}
\newcommand{\eeq}{\end{equation}}

\DeclareMathOperator{\g}{\mathfrak{g}}
\DeclareMathOperator{\h}{\mathfrak{h}}

\newcommand{\Ss}{\ensuremath{S^{3|2}}}

\newcommand{\OSPft}{\ensuremath{\text{OSP}(4|2)}}
\newcommand{\ospft}{\ensuremath{\mathfrak{osp}(4|2)}}
\newcommand{\OSPtt}{\ensuremath{\text{OSP}(3|2)}}
\newcommand{\osptt}{\ensuremath{\mathfrak{osp}(3|2)}}

\newcommand{\cS}{\mathcal{S}}

\DeclareMathOperator{\Cas}{\mathbf{Cas}}

\newcommand\bmu{\overline{\mu}}

\newcommand{\form}[2]{\ensuremath{\left( #1, #2\right)}}

\newcommand{\tb}[1]{\textbf{#1}}
\newcommand{\tbb}[1]{\overline{\textbf{#1}}}

\newcommand{\G}{G}
\renewcommand{\H}{H}

\newcommand\cP{{\cal P}}

\newcommand{\alg}[1]{\mathfrak{#1}}
\newcommand{\half}{\frac{1}{2}}

\newcommand{\Cassq}{\Cas_{\g}}

\newcommand{\ospG}{\alg{osp}(4|2)}
\newcommand{\ospH}{\alg{osp}(3|2)}

\newcommand{\abs}[1]{\lvert #1\rvert}

\DeclareFontFamily{U}{mathx}{\hyphenchar\font45}
\DeclareFontShape{U}{mathx}{m}{n}{<-> mathx10}{}
\DeclareSymbolFont{mathx}{U}{mathx}{m}{n}
\DeclareMathAccent{\widebar}{0}{mathx}{"73}

\begin{document}


\thispagestyle{empty}
\setcounter{page}{0}
\begin{flushright}\footnotesize
\texttt{DESY 14-151}\\
\vspace{2.5cm}
\end{flushright}
\setcounter{footnote}{0}

\begin{center}
{\LARGE\tb{\mathversion{bold}
On the Spectrum of Superspheres}\par}
\vspace{15mm}

{\sc  Alessandra Cagnazzo, Volker Schomerus, Vaclav Tlapak}\\[5mm]

{\it DESY Hamburg, Theory Group, \\
Notkestrasse 85, D--22607 Hamburg, Germany
}\\[5mm]

\texttt{alessandra.cagnazzo@desy.de}\\
\texttt{volker.schomerus@desy.de}\\
\texttt{vaclav.tlapak@desy.de}\\[25mm]

\tb{Abstract}\\[2mm]
\end{center}
Sigma models on coset superspaces, such as odd dimensional
superspheres, play an important role in physics and in particular
the AdS/CFT correspondence. In this work we apply recent general results
on the spectrum of coset space models and on supergroup WZNW models to
study the conformal sigma model with target space $S^{3|2}$. We construct
its vertex operators and provide explicit formulas for their anomalous
dimensions, at least to leading order in the sigma model coupling. The
results are used to revisit a non-perturbative duality between the
supersphere and the $\OSPft$ Gross-Neveu model that was conjectured by
Candu and Saleur. With the help of powerful all-loop
results for $\frac12$BPS operators in the Gross-Neveu model we are
able to recover the entire zero mode spectrum of the sigma model
at a certain finite value of the Gross-Neveu coupling. In addition,
we argue that the sigma model constraints and equations of motion
are implemented correctly in the dual Gross-Neveu description. On
the other hand, high(er) gradient operators of the sigma model are
not all accounted for. It is possible that this discrepancy is
related to an instability from high gradient operators that has
previously been observed in the context of Anderson localization.
\\


\newpage
\setcounter{page}{1}


\tableofcontents
\addtolength{\baselineskip}{5pt}

\section{Introduction}\label{sec:intro}

Non-linear sigma-models (NLSM) play an important role in physics and mathematics.
When placed on a 2-dimensional world-sheet, they give rise to renormalizable quantum
field theories \cite{Polyakov:1975rr,Brezin:1975sq,Friedan:1980jm}. Initially, 2d
NLSMs were mostly studied as toy models of 4d gauge theories, for example in order
to learn about non-perturbative features and the effect of $\theta$-terms etc.,
see e.g.\ \cite{D'Adda:1978uc}. But over the last decades, numerous direct
applications were discovered.  In string theory, for example, sigma-models
on a 2d world-sheet are the central ingredient of the perturbative definition.

The properties of sigma models depend on the choice of the target space $\mathcal{M}$
and hence on the particular problem that is addressed. Homogeneous target spaces
are particularly relevant. In these cases, the target (super)manifold $\mathcal{M}$
admits the transitive action of a continuous Lie (super)group $G$. Consequently,
$\mathcal{M}$ can be represented as the coset space $\mathcal{M} =G/H$ where $H$
is the stabilizing (super)subgroup $H \subset G$ of a point on $\mathcal{M}$.
Homogeneous (super)spaces $G/H$ for which one can find an automorphism $\gamma:
G \rightarrow G$ of order two that leaves all elements in $H \subset G$ fixed are
referred to as {\it symmetric}. Supercosets $G/H$ in which the subgroup $H$ is
fixed by an automorphism of order four  play an important role in the AdS/CFT 
correspondence, see \cite{Bena:2003wd} and references therein. While we believe 
that most of the ideas we are going to develop below apply to a wide class of 
sigma models on such generalized symmetric superspaces, our presentation and 
analysis will focus on the coset superspace OSP$(4|2)/$OSP$(3|2)$ for which 
the analysis can be made very explicit.

Understanding sigma models at strong coupling, or equivalently for
strongly curved target spaces, is of central importance. In the context
of the AdS/CFT correspondence, for example, the strongly curved regime
of the string theory is where the dual gauge theory becomes weakly
coupled. So, if we had direct control over properties of sigma models
on $AdS_{d+1}$ targets at strong coupling, it would be possible to
compare with perturbative gauge theory. Such a comparison could provide
deep new insights into the very working of gauge/string dualities, even
without the support from supersymmetry and/or integrability. On the
other hand, while we proceed to smaller values of the radius $R$, the
original background geometry dissolves and we enter a regime that is
difficult to analyse.

Nevertheless, there are a few cases in which we indeed understand sigma
models for small values of the radius. The first one that comes to mind
is the free boson compactified to a circle of radius $R$ which possesses
a description involving free fermions when $R = 1$. This can be
understood through bosonization and is known as Coleman-Mandelstam
duality \cite{Coleman:1974bu,Mandelstam:1975hb}. Sigma models on complete
intersection Calabi-Yau target spaces provide a more intricate family
of examples. For many of these models one can find so-called Gepner
points, i.e.\ values of the sigma model coupling at which the theory
possesses an exactly solvable description through certain
Wess-Zumino-Novikov-Witten (WZNW) models. This duality has been understood
most systematically through the use of linear sigma models, see
\cite{Witten:1993yc}.

In this work we present a case study for the sigma model on the
supersphere $S^{3|2}$. With this choice of the target supermanifold,
the sigma model coupling turns out to possess vanishing $\beta$ function
so that we obtain a continuous family of 2-dimensional conformal field
theories with Virasoro central charge $c=1$. We parametrize the coupling
through the radius $R$ of the bosonic base $S^{3}$. We shall discuss the
construction of vertex operators and the computation of anomalous dimensions
to leading order in $1/R$ in great detail below, thereby exemplifying
constructions and results from \cite{Candu:2013cga}. A dual description of
this supersphere sigma model has been proposed a few years ago by Candu
and Saleur \cite{Candu:2008vw,Candu:2008yw}. It involves a Gross-Neveu-like
deformation of a free field theory whose fundamental field multiplet transforms
in the fundamental representation of $\OSPft$, with four components being
fermionic and two bosonic. The free field theory corresponds to the value
$R=1$ of the radius in the supersphere sigma model. Extensive tests, mostly
based on numerical studies of a lattice discretization, have been performed
to support this proposal \cite{Candu:2008vw,Candu:2008yw}, see also
\cite{Mitev:2008yt}.

The $\OSPft$ Gross-Neveu model admits an interpretation as a current-current
deformation of an $\OSPft$ Wess-Zumino-Novikov-Witten (WZNW) model at level
$k=1$. Hence, we can exploit all-loop results from \cite{Candu:2012xc} on
anomalous dimensions of a certain subset of fields in special truly marginal
perturbations of WZNW models. The fields in question are $\frac12$BPS with
respect to the target space symmetry, or, in more mathematical terms, they
transform in maximally atypical representations. We can apply these results
to $\frac12$BPS fields in the $\OSPft$ Gross-Neveu model. For some special
value of the deformation parameter, we are able to identify the low lying
spectrum of the supersphere sigma model. The identification includes the
one-loop corrections to the conformal dimensions of the supersphere sigma
model. For sigma model fields with more than two derivatives the match
between the two models is not as convincing and we will uncover a few
discrepancies. These would have the potential to disprove the duality
conjecture of Candu and Saleur if it were not for some features of the
perturbative results on anomalous dimensions that seem to restrict their
applicability. The issue will not be settled in this work and merits a
deeper investigation.

The relation between a WZNW model at small level and a superspace sigma 
model we are about to describe illustrates several features that were 
anticipated by Polyakov in \cite{Polyakov:2005ss}. In particular, we shall 
see how singular vectors of the WZNW model are related to the sigma model
constraint and equation of motion etc. The idea to use precision data on 
deformed WZNW models in order to test non-perturbative dualities has been 
put forward previously, mostly in the context of boundary spectra, see 
\cite{Mitev:2008yt,Quella:2007sg,Candu:2009ep}. Our work is the first one
in which it is applied to bulk spectra. This is made possible mostly by 
the technical advances in \cite{Candu:2013cga,Candu:2012xc}. Let us also 
point out that conformal sigma models are not that rare when the target 
space is a supermanifold, see e.g. \cite{Bershadsky:1999hk,Berkovits:1999im,
Read:2001pz,Kagan:2005wt,Babichenko:2006uc,Candu:thesis} and references 
therein. In this sense, the ideas we put forward below should apply to 
a much wider class of examples. 

The plan of the paper is as follows. In the next section we shall review
the results of \cite{Candu:2013cga} on the construction of vertex operators
and their one-loop anomalous dimensions in coset space sigma models. These
are then worked out explicitly for the supersphere model with target space
$S^{3|2}$. For vertex operators involving a small number of derivatives we
compare our general prescriptions with more conventional constructions of
vertex operators in terms of the fundamental field multiplet. The comparison
illustrates how advantageous the new approach is in enumerating physical
fields, though once the dust settles both approaches certainly give the
same results. In section 4 we then turn to the proposed dual Gross-Neveu
model, describe its field content and the deformation away from the free
field theory. After a brief review of results from \cite{Candu:2012xc} we
analyse the low lying $\half$BPS spectrum for the value of the Gross-Neveu
coupling that is conjectured to correspond to the weakly coupled sigma
model. We shall find intriguing agreements, but also some discrepancies.
These are briefly discussed in the concluding section along with a
number of interesting open problems.

\section{The spectrum of coset sigma models}
\label{sec:spectrum_of_coset_sm}

The aim of this section is to review some results from \cite{Candu:2013cga}
concerning the spectrum of sigma models on symmetric superspaces. After a
bit of introduction we shall build a basis of fields in sigma models
on coset (super-)spaces $G/H$. At least when $G/H$ is symmetric our basis
diagonalizes the one-loop dilation operator and we can give a very simple
formula for the spectrum of one-loop anomalous dimensions. The material
of this section has been split into three different subsections of
decreasing generality. While the construction of field operators in
the first subsection holds for all coset models $\G/\H$, our discussion
of the zero modes is limited to compact $\G$. Results on the one-loop
spectrum have only been obtained for symmetric (super-)spaces, though
an extension to generalized symmetric spaces is under investigation.

\subsection{Prologue: Vertex operators for flat targets}

Before we can review what is known about the spectrum of weights we need
to recall the construction of vertex operators from \cite{Candu:2013cga}.
Let us motivate the prescription given there with a few comments on the
usual vertex operators of a free boson, i.e. a sigma model on the coset
space $S^1 =SO(2)/SO(1)$ with trivial denominator group $H = SO(1) =
\{e\}$. As is well known,
the space of such operators is spanned by
\begin{equation}
\Phi_{k;{\bf p},\overline{\bf p}} (z,\bar z) = e^{ik\theta(z,\bar z)}
{\bf p_m}(j,\partial j,\dots)\overline{\bf p}_{\bf \overline m}
(\overline \jmath,\bar \partial\overline\jmath,\dots) \ .
\end{equation}
Here, $j = j(z)$ is the current $j =
i \partial \theta$ and $\overline{\jmath}$
is of the same form but with a derivative $\bar \partial$ instead of
$\partial$, i.e.\ $\overline{\jmath}= i \bar \partial \theta$. The object
${\bf p_m}$ denotes the monomial
$$ {\bf p_m}(j,\partial j,\dots) = j^{m_1} (\partial j)^{m_2}
\cdots $$
in $j$ and its derivatives. The powers $m_i$ are components of the
multi-index ${\bf m} = (m_1,m_2,\dots)$ we have placed on ${\bf p}$.
Of course, the definition of $\overline{\bf p}$ is similar, but with
derivatives $\bar\partial$ instead of $\partial$. Note that the
multi-index $\tbb{m}$ is independent of ${\bf m}$.

The operators $\exp(ik\theta)$ are associated to the zero modes of the free
boson, i.e.\ there is one such operator for each function on the target space.
For ${\bf m} =0 = \tbb{m}$ we obtain the usual tachyon vertex operators.
The choice ${\bf m}=(1,0,0,\dots)= \tbb{m}$ corresponds to the vertex
operators for massless states etc.

\subsection{Vertex operators for \texorpdfstring{$\G/\H$}{G/H}}
\label{Subsec:Vertex_operators}

In generalizing this discussion to non-trivial coset models $G/H$ we
must address how to replace currents $j$ and $ \overline\jmath$, the
{\it tail} monomials ${\bf p}$ and $\overline{\bf p}$ and the {\it zero
mode} contributions $\exp(ikX)$.

Let us begin with the fields $j$ and $\overline\jmath$. One could imagine
to simply take derivatives of coordinate fields $\theta_J$ that are associated
with some choice of coordinates on $G/H$. While this works just fine for
a flat target space, it is not the smartest choice for curved
backgrounds. Instead, we shall adopt the definition
\begin{equation}
\label{eq:jdef}
j_\alpha := E_\alpha^J(\theta) \partial \theta_J \quad ,\quad
\overline \jmath_{\alpha} := E_\alpha^J (\theta) \bar \partial \theta_J \
\end{equation}
where $E_\alpha^J$ is the vielbein for our coset space. Equivalently, if
we think of the points on $G/H$ as being parametrized by orbits of group
elements $g \in G$ under the right action of $H$, we can also construct
$j$ and $\overline \jmath$ as
\begin{equation}
\label{eq:defcurrents1}
j_{\alpha} = \form{ g^{-1} \partial g}{t_\alpha}\ , \qquad
\overline{\jmath}_\alpha = \form{g^{-1} \bar \partial g}{t_\alpha}\ .
\end{equation}
Here, $t_\alpha$ runs through a basis in the quotient space $\mathfrak{m}=
\mathfrak{g}/\mathfrak{h}$. The space $\mathfrak{m}$ carries an action of
the denominator Lie (super-)algebra $\mathfrak{h}$. Its dimension coincides
with the dimension of $G/H$. Note that there is one crucial difference with
respect to the flat target $S^1$, namely our fields $j$ and $\overline\jmath$
transform non-trivially under the action of the denominator algebra. Of course,
physical fields of the coset model must be invariant. Hence, it will be important
to keep track of how the composite fields we are about to construct transform
under $\mathfrak{h}$.

A field can contain arbitrary products of $j_{\alpha}$ and $\overline{\jmath}_{\alpha}$
and their derivatives, just as for flat targets. Since the multiplets $(j_{\alpha})$
and $(\overline{\jmath}_{\alpha})$ transform in the representation $\mathfrak{m}$ of
$\h$, we can build tails in any subrepresentation $[\mu]$ that appears in some tensor
power of  $\mathfrak{m}$. More precisely, we can pick two multi-indices ${\bf m}$ and
$\tbb{m}$ as in our discussion of the compactified free boson and then choose two
intertwiners
\begin{equation} \label{PbP}
 {\bf P}_{\mu,{\bf m}}: \bigotimes_{i} \mathfrak{m}^{\odot m_i}
       \rightarrow [\mu] \ \quad ,\quad
 \overline{\bf P}_{\bmu,\tbb{m}}: \bigotimes_{i} \mathfrak{m}^{\odot{\bar m_i}}
       \rightarrow [\bmu]\ .
\end{equation}
Here, we used $\mathfrak{m}^{\odot m}$ to denote the m-fold (graded) symmetric
tensor power of $\mathfrak{m}$. Given any such intertwiner, we construct
the tail factor
\begin{equation}
 \label{eq:defproductcurrents1}
{\bf P}_{\mu,\bf m}(j,\partial j,\dots) = {\bf P}_{\mu,{\bf m}} \left[ j^{\otimes^{m_1}}
\otimes (\partial j)^{\otimes^{m_2}} \otimes \cdots \right]
\end{equation}
and similarly for the second contribution that involves $\overline\jmath$ and
its derivatives with respect to $\bar \partial$. We have used tensor products and
powers instead of ordinary ones to remind us that $j$ is a multi-component object.
Note that there is a finite number of intertwiners ${\bf P}_{\mu,{\bf m}}$ and
$\overline{{\bf P}}_{\bmu,\tbb{m}}$ for any given choice of ${\bf m}$ and
$\tbb{m}$. This finite choice has no analogue in a flat background.

Having discussed the tail of our vertex operators, we also need to
address the zero mode factors. In the compactified free boson the zero mode contribution
was a function on the target space. Functions on the coset space $\G/\H$ can be
thought of as $\H$-invariant functions on the group $\G$. But since our tail
factors transform non-trivially under $\H$, it seems natural to admit zero mode
contributions whose transformation behavior under the right action of $\H$ on $\G$ is
non-trivial as well. More precisely, for any given representation $S_\lambda$
of $\H$ on the carrier space $\cS_\lambda$ let us consider the following
space of $\cS_\lambda$-valued functions on $\G$,
\begin{equation}
\Gamma_\lambda = \Gamma_\lambda(\G/\H) = \{ F \in L_2(G) \otimes \cS_\lambda  :
F(gh) = S_\lambda(h^{-1}) F(g)\
\forall h\in \H\}\ .
\end{equation}
Elements of the linear space $\Gamma_\lambda$ may be considered as sections in
a homogeneous vector bundle on $G/H$ \cite{Bott}. We will analyse the structure 
of these vector bundles in the next subsection.

At this point we have discussed three ingredients of our vertex operators,
namely the tail factors ${\bf P}_{\mu,{\bf m}}$ and $\overline{{\bf P}}_{\bmu,\tbb{m}}$
along with the zero mode contribution $V \in \Gamma_\lambda$. These transform
in the representations $\mu,\bmu$ and $\lambda$ of  the denominator algebra
$\mathfrak{h}$. Obviously, a physical field in the coset model must be
$\mathfrak{h}$ invariant. Hence, we must glue our three ingredients with
an intertwiner
\begin{equation}\label{C}
 {\bf C}^{\lambda\mu\bmu}: [\lambda]\otimes [\mu]
\otimes [\bmu]  \ \rightarrow\ \mathbb{C}
\end{equation}
from the triple tensor product between the representations $[\lambda],
[\mu]$ and $[\bmu]$ of the denominator algebra $\h$ to the
complex numbers. Fields of the coset model now take the form
\beq
\label{eq:fieldscosetmodel}
 \Phi_{\Lambda,\lambda,\mu,\bmu}(z,\bar z) =
V_{\Lambda\lambda}(z,\bar z)\,
{\bf P}_{\mu;\tb{m}}(j,\partial j,\dots) \,
\overline{\bf P}_{\bmu;\tbb{m}}(\overline\jmath,
\bar\partial \overline\jmath,\dots)\, {\bf C}^{\lambda\mu\bmu}\ ,
\eeq
where $V_{\Lambda\lambda} \in \Gamma_\lambda$
is a section that transforms in the representation $\Lambda$ of
the numerator algebra $\g$. By construction, these fields are invariant
under the action of the denominator group $H$. On the other hand, the
action of the numerator group $G$ is non-trivial. It is determined by the
way the section $V_{\Lambda\lambda}$ transforms. The label $\Lambda$
is the curved space analogue of the linear momentum $k$ in a circular
target $S^1$.

The labels $(\Lambda,\lambda,\mu,\bmu)$ we have placed on
the symbol $\Phi$ do not keep track of all the freedom we have in the
construction of vertex operators. In order to count all possible fields
of the coset model one needs to count the intertwiners ${\bf P},
\overline{\bf P}$ and ${\bf C}$ that were introduced in eqs.\ \eqref{PbP}
and \eqref{C}, respectively. In addition, there is often some freedom in
the choice of the section $V_{\Lambda\lambda} \in \Gamma_\lambda$.
While the number of intertwiners may be determined straightforwardly
from the fusion rules of the Lie (super-)algebra $\mathfrak{h}$, the
space of sections in homogeneous vector bundles requires input from
harmonic analysis. We will analyse the space $\Gamma_\lambda$ in the
next subsection. For $O(N)$ vector models, i.e.\ the coset sigma models
with target space $O(N)/O(N-1)$, the space of fields has been counted
in \cite{Candu:2013cga} and the result was shown to agree with other
descriptions of the field space for these models.

\subsection{Homogeneous vector bundles on \texorpdfstring{$\G/\H$}{G/H}}
\label{sec:homo_vec_bundles}

As we explained in the previous subsection, a good control over vertex
operators of coset models requires some knowledge about sections in
homogeneous vector bundles over $\G/\H$  and their transformation
behavior under the (left) action of $\G$. Our main goal in this subsection
is to explain the decomposition
\begin{equation} \label{Gammaexp}
\Gamma_\lambda \cong \sum_\Lambda n_{\Lambda\lambda}\  [\Lambda]\ .
\end{equation}
Here, the linear space $\Gamma_\lambda$ is considered as a representation
of the numerator Lie (super-)algebra $\mathfrak{g}$. The summation on the right
hand side runs over irreducible representations $[\Lambda]$ of this algebra.
Let us stress that for Lie superalgebras, the sum is not direct, at least not
in general. We will return to this issue below.

In the expansion \eqref{Gammaexp}, each summand $[\Lambda]$ appears with
some multiplicity $n_{\Lambda\lambda}$. Following standard mathematical
notation, we shall also write
\begin{equation}
 n_{\Lambda \lambda} = [\Gamma_\lambda : \cS_\Lambda]  
\end{equation}
for the number of times a given irreducible representation $\cS_\Lambda$
of $\mathfrak{g}$ appears in (the decomposition series of) the space $\Gamma_\lambda$
of sections. It is a central result from harmonic analysis of compact supergroups that
\begin{equation} \label{fundlemma}
[\, \Gamma_\lambda\, :\, \cS_\Lambda\, ] =
[\, \cP_\Lambda|_\mathfrak{h}\, :\, \cP_\lambda\, ] \ .
\end{equation}
The objects $\cP_\Lambda$ and $\cP_\lambda$ denote representations
of the Lie superalgebras $\g$ and $\h$, respectively. These particular
representations are called projective covers, see e.g. \cite{Germoni2000:MR1840448,
Quella:2013rev} for a precise definition and more background. They coincide 
with the irreducible representations $\cS_\Lambda$ and $\cS_\lambda$ when no
shorting conditions are satisfied, i.e. when both $\Lambda$ and $\lambda$
are non-BPS. The case of BPS (or atypical) multiplets will be discussed
in more detail below. After restriction to $\h \subset \g$, the
representation $\cP_\Lambda$ gives rise to a representation
$\cP_\Lambda|_{\h}$ of ${\h}$. The number on the right
hand side of equation \eqref{fundlemma} denotes the number of 
times the representation $\cP_\lambda$ appears in the representation
$\cP_\Lambda|_{\h}$.

All this might seem a bit abstract at first. So, let us briefly
illustrate the content of  eq.\ \eqref{fundlemma} for the coset
space $S^2 = $ SU$(2)/$U$(1)$. In this case, there exists an infinite
set of complex line bundles which are parametrized by the
monopole number $k \in \mathbb{Z}$. This number and hence the
associated bundles are in one-to-one correspondence with
irreducible representations $\cS_k$ of the denominator group
$H$ = U$(1)$. For monopole number $k= 0$ we are dealing with
the trivial line bundle, i.e.\ with functions on $S^2$. Of
course we know very well how the space of functions decomposes
under the action of $\mathfrak{su}(2)$: Each integer spin representation
appears with multiplicity one. We may recover this fact from
our formula \eqref{fundlemma} as follows. The space of functions
on $S^2$ is associated to the label $\lambda = 0$. We want to
know how many times an irreducible representation $\cS_\Lambda
= \cS_j$ of $\mathfrak{su}(2)$ appears in the decomposition of $\Gamma_0$.
According to eq.\ \eqref{fundlemma}, this number is given by
\begin{equation}[\Gamma_0: \cS_j] = [\cS_j|_{\text{U}(1)}:\cS_0] =
\left\{ \begin{array}{ll}  1 & \text{for }\  j \in \mathbb{N}\\[2mm]
                           0 & \text{for } \ j \in \mathbb{N}+\frac12
                           \end{array} \right. . 
\end{equation}
Here ${\mathcal S}_0$ denotes the trivial representation of $\h$.
For bosonic Lie groups, we do not have to distinguish between
projective covers $\cP_j$ and irreducibles, i.e.\ $\cS_j = \cP_j$. The
second equality follows from the fact that the spin $j$
representation $\cS_j$ contains exactly one state on which the
generator $J^3$ of the $\mathfrak{u}(1)\subset \mathfrak{su}(2)$ has zero
eigenvalue if and only if $j$ is integer. For non-trivial monopole line
bundles, the evaluation proceeds along the same lines. In this
case the space $\Gamma_k$ of sections contains each integer
spin representation $\cS_j$ satisfying $j\geq k$ with multiplicity
one.

The only additional complication we have to deal with in
applying eq.\ \eqref{fundlemma} to superspaces comes from the
distinction between irreducibles and projective covers. For
typical (long) multiplets $\cS_\Lambda$ of a Lie superalgebra
$\g$, the projective cover $\cP_\Lambda$ agrees with $\cS_\Lambda
= \cP_\Lambda$. But if $\cS_\Lambda$ is an atypical (short)
multiplet then $\cP_\Lambda \neq \cS_\Lambda$ is an indecomposable
representation. It should be considered as a very specific
`composite' representation that is built from several short
multiplets. For the Lie superalgebra $\mathfrak{g} = \ospft$
the projective covers are discussed explicitly in appendix
\ref{app:bg_osp42}. Of course, short representations of the denominator
algebra $\h$ can also be combined into projective covers, see appendix
\ref{app:osp32} where the projective covers for $\mathfrak{osp}(3|2)$
are discussed. Let us finally mention that upon restriction from
$\mathfrak{g}$ to the subalgebra $\h \subset \g$, a projective cover
$\cP_\Lambda$ decomposes into a direct sum of projective covers
$\cP_\lambda$. Hence, the numbers on the right hand side of eq.\
\eqref{fundlemma} are well defined. We shall compute them
for homogeneous vector bundles on the supersphere $S^{3|2}$ later on.

Let us briefly mention one simple example that can be used to
illustrate how important the distinction between irreducibles and
projective covers is. To this end we consider the homogeneous
vector bundle $\Gamma_{\text{ad}}$ on the supersphere \Ss\ that is
associated with the adjoint representation of the denominator
algebra $\mathfrak{osp}(3|2)$. It turns out that this bundle contains
two multiplets of sections which transform in the adjoint representation
$\cS_\text{Ad}$ of the numerator algebra $\mathfrak{osp}(4|2)$, i.e.\
$[\Gamma_{\text{ad}}:\cS_\text{Ad}]=2$.
On the other hand, the adjoint representation of $\mathfrak{osp}(4|2)$
is 17-dimensional and that of $\mathfrak{osp}(3|2)$ is 12-dimensional.
Hence, for dimensional reasons, the restriction of $\cS_\text{Ad}$ to
$\mathfrak{osp}(3|2)$ contains $\cS_\text{ad}$ only once,
\begin{equation}
2 = [\cP_\text{Ad}|_\mathfrak{h}, \cP_\text{ad}] \neq
[\cS_\text{Ad}|_\mathfrak{h},\cS_\text{ad}]=1.
\end{equation}
This example demonstrates that harmonic analysis on superspaces
requires a bit of extra care precisely because of the existence of
BPS representations.

Before we conclude this subsection let us stress once more that formula
\eqref{fundlemma} is restricted  to compact (super-)algebras. This does not mean
that similar control of homogeneous vector bundles can not be achieved when $\G$ is
non-compact. As long as $\H$ is compact, one can continue to derive results
on the decomposition of homogeneous vector bundles from the harmonic analysis
of $\G$. So, if the latter is understood, homogeneous vector bundles pose no
additional problems. When $\H$ in non-compact, however, normalizable sections
of on $\G/\H$ are no longer obtained from normalizable functions on $\G$ and
hence cosets  with non-compact denominator require an independent analysis.
Nevertheless, the decomposition of homogeneous vector bundles is known in
many concrete examples.

\subsection{One-loop anomalous dimensions}

While our construction of vertex operators in coset sigma models was completely
general and the property \eqref{fundlemma} holds for all homogeneous vector bundles
on quotients $\G/\H$ of a compact Lie (super-)group $\G$, the following results on
the one-loop corrections to the spectrum of coset models have only been derived for
symmetric (super-)spaces, although work on so-called generalized symmetric spaces,
including those relevant for the AdS/CFT correspondence, is in progress.

The computations carried out in \cite{Candu:2013cga} show that the one-loop anomalous
dimensions depend only on the representation labels $\Lambda,\lambda,\mu,\bmu$ and
not on the intertwiners ${\bf P},\overline{{\bf P}}$ and ${\bf C}$ that enter the
construction of fields \eqref{eq:fieldscosetmodel} in the coset model. This is why
we labeled our fields $\Phi$ by a subscript that makes no reference to the precise
choice of intertwiners.

At zero sigma model coupling, i.e.\ for $R = \infty$, the sigma model fields possess
their naive dimensions $(h_\infty,\bar h_\infty)$ that are given by the number of
derivatives,
\begin{equation}
  h_\infty = \sum_{j=1} j\, m_j \quad , \quad
   \bar h_\infty = \sum_{j=1} j\,  \overline{m}_j \ . 
\end{equation}
Once we turn on the interaction, these scaling weights are shifted by
the so called anomalous dimension $\delta_R h$, i.e.\ at some finite
value of the coupling $R$ the scaling weights have the form
 
\begin{equation}\bigl(h(R),\bar h(R)\bigr) = (h_\infty + \delta_R h,
    \bar h_\infty + \delta_R h)\ . 
\end{equation}
According to \cite{Candu:2013cga}, the leading contribution to the anomalous
dimension takes the form
\begin{equation}\label{eq:oneloopdimension}
	\delta^{(1)}_{R} h =  \frac{1}{2 R^2} \bigl(\Cas_{\g}(\Lambda)
    - \Cas_{\h}(\mu) - \Cas_{\h}(\bmu)\bigr)\ .
\end{equation}
In the derivation the result actually emerges as a sum of two different
pieces that are associated with the zero mode factor and the tail of the
vertex operator, respectively. Recall that the zero mode factor $V_{\Lambda
\lambda}$ is a section in a homogeneous vector bundle $\Gamma_\lambda$. Such
sections are acted upon by the Lichnerowicz Laplacian $\Delta_L$, whose
eigenvalues were expressed through the quadratic Casimir operators of
$\mathfrak{g}$ and $\mathfrak{h}$ in \cite{Pilch:1984xx},
\begin{equation} \Delta_L V_{\Lambda\lambda}(\theta) =
\bigl(\Cas_{\g}(\Lambda) - \Cas_{\h}(\lambda)
\bigr) V_{\Lambda\lambda}(\theta)\ .
\end{equation}
The contribution of the tail factors to the annomalous dimension can
be written as a spin-spin interaction between fields $j$ and
$\overline{\jmath}$. It leads to a term of the form $\Cas_{\h}(\lambda)-
\Cas_{\h}(\mu) - \Cas_{\h}(\bmu)$. Note that the first term in this
combination cancels the constant shift $\Cas_{\h}(\lambda)$ in the
eigenvalues of the Lichnerowicz Laplacian so that we end up with the
expression given in eq.\ \eqref{eq:oneloopdimension}.

Formula \eqref{eq:oneloopdimension} is actually very general. It
holds for {\it all} sigma models on symmetric superspaces with vanishing
beta function. When properly interpreted, see \cite{Candu:2013cga}, it can
also be used for models with world-sheet supersymmetry, such as e.g. the
$N=2$ worldsheet supersymmetric sigma model on complex projective superspace
$\mathbb{C}\text{P}^{3|4}$ etc. In applications to non-conformal theories, such as e.g.\
the usual O$(N)$ models, the formula for $\delta^{(1)} h$ requires a
simple additional term,
\begin{equation}\label{eq:oneloopdimensionnp}
\delta^{(1)}_R h =  \frac{1}{2 R^2} \Bigl(\Cas_{\g}(\Lambda)
    - \Cas_{\h}(\mu) - \Cas_{\h}(\bmu) + \Cas_{\h}(\mathfrak{m})\,
    \sum_i(m_i +\overline m_i) \Bigr)\ .
\end{equation}
Since vanishing of the one-loop beta function requires that
$\Cas_{\h}(\mathfrak{m})=0$ we recover the formula \eqref{eq:oneloopdimension}
for conformal sigma models. Our simple formula \eqref{eq:oneloopdimension} or
rather its generalization \eqref{eq:oneloopdimensionnp} summarizes and extends
the results of many papers in which anomalous dimensions, mostly dealing with
$\mathfrak{g}$-invariant fields, have been studied model by model, see e.g.
\cite{Kravtsov:1989qs,Wegner:1990AD,Wegner:1991gf,Mall:1993wr,Vasiliev:1993ux,
Lang:1993ct,Derkachov:1997qv}. That all these computations may
be captured by a single universal formula \eqref{eq:oneloopdimensionnp} is
quite remarkable. Of course, this success is intimately tied to the
construction \eqref{eq:fieldscosetmodel} of vertex operators. We now see how
well this construction was adapted to the computation of 1-loop anomalous
dimensions.

Much of the previous work on anomalous dimensions of (high-)gradient
operators in sigma models was motivated by a somewhat puzzling instability
that has first been observed in O$(N)$ vector models \cite{Kravtsov:1989qs} and
later understood to be a rather generic feature of sigma model perturbation
theory, see \cite{Ryu:2010iq} and references therein. This instability
arises because of the negative sign in front of the terms $\Cas_{\h}(\mu)$
and $\Cas_{\h}(\bmu)$. Naively one might think that high gradient
operators, i.e. operators \eqref{eq:fieldscosetmodel} for which
$\sum_j j (m_j+ \overline{m}_j)=h_\infty+\bar h_\infty$ is large, are highly
irrelevant. But it turns out that some of these operators acquire a very large
negative anomalous dimension. More precisely, one can show that for every choice
of the sigma model coupling $R^{-2}$, no matter how small, one can find
a $\mathfrak{g}$-invariant high gradient operator ${\cal O} = \Phi_{0,\lambda,
\mu,\bmu}$ such that
 \begin{equation} h_\infty({\cal O}) + \bar h_\infty({\cal O})
- \frac{1}{R^2} \bigl(\Cas_{\h}(\mu)+ \Cas_{\h}(\bmu)\bigr) < 2  \ .
\end{equation}
This is because $\Cas_{\h}(\mu)$ grows quadratically with the weights of
the representation $\mu$ and the maximal weight grows linearly with the
number of currents $j$ in the tail. On the other hand, the positive
contribution $h_\infty({\cal O})$ only grows linearly in the number of
$j$s. The argument shows that (infinitely many) high gradient operators
 become relevant for arbitrarily small sigma model coupling. One could have
hoped that higher orders in perturbation theory correct the issue, but
they turn out to make things even worse \cite{Castilla:1996qn}. We would
be ready to conclude that sigma models are inherently unstable if it were
not for the many independent studies, e.g. through lattice discretizations,
that display no pathologies. As far as we know, the problem has never been
resolved but it is something to be kept in mind as we proceed.

\section{The spectrum of the supersphere \texorpdfstring{$S^{3|2}$}{S32}}
\label{sec:supersphere_spectrum}

The aim of this section is twofold. Partly, we would like to illustrate the
general results we have reviewed in the previous section through the simplest
nontrivial example of an interacting conformal sigma model, namely the theory
with target space $S^{3|2}$. This supersphere can be considered as
a quotient $\G/\H$ of the compact supergroup $\G = \text{OSP}(4|2)$ by the
subgroup $\H = \text{OSP}(3|2)$. Since the latter is fixed by an order two
automorphism of the former, the supersphere $S^{3|2}$ is a compact
symmetric superspace. Hence, all the results we outlined in
the previous section
apply to this case. Our task is to work them out explicitly. This will
require some input from the representation theory of $\mathfrak{osp}(4|2)$
and $\mathfrak{osp}(3|2)$ which can be found in several appendices. The
second purpose of this section is to gather some data about the supersphere
sigma model that we can later use to test the conjectured duality with the
OSP$(4|2)$ Gross-Neveu model.

We will begin by describing several equivalent formulations
of the supersphere sigma model. Concrete results on low gradient operators
and their anomalous dimensions are worked out in the second subsection.
In the third subsection we describe the more conventional construction of
(low gradient) vertex operators in terms of the fundamental field of the
non-linear sigma model. While this turns out to be significantly more
cumbersome than the approach advocated in the previous subsection, it will
allow us to understand the impact of symmetries and equations of motion.

\subsection{The supersphere sigma model}

The most basic description of the supersphere $S^{3|2}$ is as a co-dimension
one supermanifold in the flat superspace $\mathbb{R}^{4|2}$ defined by the
equation
\begin{equation}\label{constraint}
X \cdot X := \sum_{j=1}^{4}  x^2_j + 2 \eta_{1} \eta_{2}
\ = \ 1\ \ .
\end{equation}
Here, $x_j, j = 1, \dots,4,$ and $\eta_1, \eta_2,$ are the bosonic and fermionic
coordinates of $\mathbb{R}^{4|2}$, respectively. We shall often combine these
coordinates into a multiplet of supercoordinates $X = (X_A) = (x_j,\eta_1,
\eta_2)$. For a pair $X$ and $Y$ is such multiplets the inner product $\cdot$
is defined as
 \begin{equation} X \cdot Y = \sum_jx_j y_j + \eta_1\xi_2 - \eta_2 \xi_1 \ .  
\end{equation}
Here, we have denoted the fermionic coordinates of $Y$ by $\xi_1$ and $\xi_2$.
We can now write the action of the associated sigma model as
\begin{equation}\label{SSSmodel}
{\cal S}^{\text{SM}}[X,\rho] = \frac{R^2}{2\pi} \int d^2z \left(
\partial X \cdot \partial X - \rho X \cdot X  \right) \ .
\end{equation}
Here $\rho$ is a Lagrange multiplier that implements the supersphere constraint
\eqref{constraint}. The parameter $R$ can be interpreted as the radius of the
supersphere. In the regime where $R$ is large, the sigma model is weakly coupled
and perturbation theory should give reliable results. The equations of motion
for the field multiplet $X$ read
\begin{equation} \label{eom}
 \partial \bar \partial X = (\partial X \cdot \bar \partial X) X\ .
\end{equation}
From our description of the supersphere through equation \eqref{constraint}
it is evident that \Ss\ comes equipped with an \ospft\ action. In fact, the
Lie superalgebra \ospft\ acts on the embedding space $\mathbb{R}^{4|2}$ through
its fundamental representation. By the very definition of \ospft\ this action
respects the constraint \eqref{constraint}. The supersphere \Ss\ can be obtained
from the supergroup $\OSPft$ by taking the following quotient
\begin{equation}\label{coset}
  \Ss \ = \ \OSPft/\OSPtt  \
\end{equation}
with respect to the right action of the subsupergroup $\OSPtt \subset \OSPft$.
The latter appears as the stabilizer of a point $X = (X_A) = (1,0,0,\dots)$ on
the supersphere. Since this stabilizer is left invariant by the reflection of
the first coordinate, the quotient \eqref{coset} is a symmetric superspace. In
conclusion, we have shown that the sigma model \eqref{SSSmodel} possesses all
the properties that we assumed in the previous section.

In order to get a better feeling for how non-trivial the supersphere sigma model
really is, we solve the constraint~\eqref{constraint} explicitly. To this end, we
parametrize \Ss\ through three angular coordinates $\vartheta_j$ and $2$ fermionic
variables $\eta_b$. The line element takes the following form
\begin{equation}\label{metric}
ds^2 \ = \ 2  (1-\eta_1\eta_2) d\eta_1d\eta_2 +  (1-2 \eta_1\eta_2)d\Omega_3
\end{equation}
where
$$
d\Omega_3 \ = \ d\vartheta^2_1 + \cos^2 \vartheta_1\ d\vartheta_2^2 +
\sin^2 \vartheta_1 \ d\vartheta_3^2
$$
is the usual line element of the 3-dimensional unit sphere. In the sigma model,
the coordinates are promoted to fields and the action reads
\begin{equation} \label{PCM}
\begin{split}
{\cal S}^{\text{SM}}[\vartheta,\eta] \ = \ & \frac{R^2}{2\pi} \int d^2z
\bigl(
 2 (1-\eta_1\eta_2) (\partial\eta_1 \bar \partial \eta_2
 - \partial \eta_2 \bar\partial \eta_1)\phantom{\frac12} 
 \\[2mm]
 & \hspace*{1cm}
 \phantom{\frac12}
+ (1-2\eta_1\eta_2) (\partial \vartheta_1 \bar \partial
\vartheta_1 + \cos^2 \vartheta_1\, \partial \vartheta_2 \bar
\partial\vartheta_2 + \sin^2
\vartheta_1\, \partial \vartheta_3 \bar \partial\vartheta_3)\bigr)\ . \\
\end{split}
\end{equation}
For the sigma model on the purely bosonic 3-sphere the coupling $R$ runs and
in order for the flow to end in a non-trivial fixed-point one must add a Wess-Zumino
term \cite{Witten:1983ar}. But the presence of the two fermionic directions changes
the situation profoundly. As shown in \cite{Read:2001pz}, the $\beta$-function of
the sigma model on \Ss\ is the same as for a bosonic sigma model on a sphere $S^d$
whose dimension $d=3-2=1$ is given by the difference between the number of bosonic
and fermionic coordinates. Consequently, the $\beta$-function vanishes for the
sigma model on \Ss, i.e.\ the model \eqref{PCM}, defines a family of non-unitary
interacting conformal field theories at central charge $c=1$ with continuously
varying exponents.

Before we apply the results reviewed in the previous section to this model
let us note that the action \eqref{PCM} can be written very compactly if we
factorize the metric with the help of the super-Vielbeins $E^J_\alpha(\vartheta,
\eta)$,
\begin{equation}
g^{IJ}(\vartheta,\eta) \ :=\  \kappa^{\alpha \beta}
E^I_\alpha(\vartheta,\eta) E^J_\beta(\vartheta,\eta) (-1)^{|\beta|(|I|+|\alpha|)}
\end{equation}
where $\kappa$ is the invariant form of \ospft\ and the indices
$\alpha,\beta$ run over directions along the quotient $\mathfrak{m} =
\ospft/\osptt$. We can now
combine the Vielbeins with the derivatives of the coordinate fields $(\theta_J)
= (\vartheta_j,\eta_a)$ as in eq.\ \eqref{eq:jdef} to obtain
\begin{equation}
 {\cal S}^{\text{SM}}[\theta]  =   \frac{R^2}{2\pi} \int d^2z\, g^{IJ}(\vartheta,\eta)
 \, \partial \theta_I \bar \partial\theta_J =  \frac{R^2}{2\pi} \int d^2z
 \, \kappa^{\alpha\beta}
  j_\alpha(z,\bar z) \overline{\jmath}_\beta(z, \bar z) \ .
\end{equation}
Of course, all the non-linearity of the action \eqref{PCM} is just hidden
in the complicated structure of the fields $j$ and $\overline{\jmath}$.
Note that the latter transform in the fundamental representation of the
stabilizer subgroup $\OSPtt$. In the action the corresponding index
$\alpha$ is contracted with the $\beta$ so as to give an invariant.

Unlike the sigma model on $S^1 =$ U(1), the theory defined by
the action \eqref{PCM} is not free. For large radius $R$, the
model is weakly coupled and its properties may by studied
perturbatively. But as we pass to a more strongly curved
background, computing quantities as a function of the radius $R$
may seem like a very daunting task. This is even more so because
there is very little symmetry to work with. As a conformal field
theory, the sigma model on the supersphere possesses the usual chiral
Virasoro symmetries. But for a model with multiple bosonic
coordinates the two sets of chiral Virasoro generators are not
sufficient to make the theory rational. Since there are no
efficient tools to construct the theory at generic values of the
radius parameter $R$, finding a dual description whose perturbative
regime describes a strongly curved supersphere is of obvious
interest.

\subsection{Vertex operators and anomalous dimensions}
\label{sec:vertex_ops_and_anom_dim}

Before we can begin constructing vertex operators for the supersphere
sigma model we need a little bit of background on representations of
both $\ospft$ and $\osptt$. A much more comprehensive discussion can
be found in the appendices. It is heavily based on two papers by 
van der Jeugt \cite{VanDerJeugt:1985hq,VanderJeugt:1984zs}.  

The Lie superalgebra $\ospft$  possesses
the bosonic subalgebra $\mathfrak{so}(4) \oplus \mathfrak{sp}(2)$.
Since this has rank $r=3$, generic representations are labeled by
triples of weights $[j_1,j_2,j_3]$. Atypical (or BPS) representations
satisfy a single shortening condition. The possible conditions are
listed in eq. \eqref{atypicality conditions}. With one such condition
relating the three weights $j_i$, atypical representations $\Lambda_{l,k}$
are labeled by two integers $l \geq 0$ and $k$. The precise relation
between $l,k$ and the weights $j_i$ are given in eqs.\ \eqref{atyp1}
and \eqref{atyp2}. Let us only note that the label of the trivial
representation is $\Lambda_{0,0}$ while that of the 17-dimensional
adjoint is $\Lambda_{0,1}$. The representations $\Lambda_{l,0}$ on
the other hand are associated with (graded) symmetric traceless
tensors of  $\ospft$.

In the atypical representation $\Lambda_{l,k}$,  the quadratic Casimir
element $\Cas_{\g}$ takes the value
 \begin{equation} \Cas_{\g}(\Lambda_{l,k}) = l^2 \ .  
\end{equation}
We conclude that the Casimir element $\Cas_{\g}$ is insensitive to
the second label $k$ of $\Lambda_{l,k}$. Atypical representations with
the same value of the Casimir element are said to belong to the same
block. Representations from the  same block may appear
within larger indecomposables, in particular they make up the projective
covers $\mathcal{P}_{\Lambda_{l,k}}$. The composition series of these
indecomposables are given in eqs. \ \eqref{projdec1}-\eqref{projdec4}.

Let us turn our attention to the Lie superalgebra $\osptt$. In this
case, the bosonic subalgebra $\mathfrak{so}(3) \oplus \mathfrak{sp}(2)$
has rank two and hence generic representations are labeled by a pair
$[q,p]$ of weights. The atypicals $\lambda_0$ and $\lambda_q =
[q,2q-1]$, $q \geq 1/2,$ form a 1-parameter family of representations
that satisfy a single shortening condition. The label $\lambda_0$
is reserved for the trivial representation, $\lambda_{1/2}$ is the
5-dimensional fundamental. In the case of $\osptt$, the adjoint is
not atypical.  Its label is $\lambda_\text{ad}= [1,0]$.

In the representation $[q,p]$ the quadratic Casimir element $\Cas_{\h}$
of $\osptt$ takes the values
\begin{equation} \Cas_{\h} \bigl([q,p]\bigr) = (p+2q)(p-2q+1) \ .  
\end{equation}
We see that it vanishes for atypicals $\lambda_q$. All these atypicals
belong to the unique single block from which indecomposables can be
built. Once again, the most relevant indecomposables are the projective
covers ${\cal P}_{\lambda_q}$ of atypicals. Their composition series
are displayed in eqs.\ \eqref{projdec1}-\eqref{projdec4}.

With these notations set up we can begin to construct vertex operators. Our
goal is to find all vertex operators with up to two derivatives that transform
in $\frac12$BPS representations $\Lambda_{l,k}$ of $\ospft$. Let us start with
the zero modes. By definition, these fields have vanishing scaling dimension at
$R=\infty$ so  they cannot contain any currents $j$ or $\overline{\jmath}$.
Consequently, the $\osptt$ representations $\mu,\bmu$ and $\lambda$ that label
our vertex operators \eqref{eq:fieldscosetmodel} are all trivial. Thus,
the head must  be taken from
\begin{equation}\label{Gamma0}
\Gamma_0 = \Gamma_{\lambda_0} =  \bigoplus_{l=0}^\infty \Lambda_{l,0}
\ ,
\end{equation}
where $\Lambda_{l,0} = \frac12[l+1,l-1,-l-1]$
for $l>0$ and $\Lambda_{0,0}$ is the trivial representation. In order to find
the decomposition displayed on the right hand side, we employed the decomposition
formulas \eqref{pcdec1}-\eqref{pcdec3} along with the fundamental results
\eqref{fundlemma}. The summation is over all those representations
$\Lambda$ of $\ospft$ for which the restriction of ${\mathcal{P}}_\Lambda$ to the
subalgebra $\osptt$ contains $\mathcal{P}_{\lambda_0}$. Our formulas in appendix
\ref{app:osp42_osp32} only list the decompositions for atypical representations
$\Lambda = \Lambda_{l,k}$ but it is not difficult to see that typical (long)
multiplets never contain $\mathcal{P}_{\lambda_0}$ in their decomposition. Hence,
the formula \eqref{Gamma0} is exact, i.e.\ it accounts  for all elements of
$\Gamma_0$ not just for those that transform in $\frac12$BPS representations.
Of course, the space $\Gamma_0$ is nothing but the space of functions on the
supersphere $\Ss$. Aside from the trivial representation $\Lambda_{0,0}$ of
$\ospG$, which has vanishing Casimir, all other operators acquire a non-zero
anomalous dimension,
\begin{equation}\label{1loopzeromodes}
\delta^{(1)}_R h(V_{\Lambda_{l,0},\lambda_0}) = \frac{1}{2 R^2} \Cas_{\g}(\Lambda_{l,0}) =
\frac{l^2}{2 R^2}\ .
\end{equation}
The next set of operators we would like to look at are the operators of weight
$(h_\infty,\bar h_\infty)=(1,0)$. Such operators contain a current $j$ and hence  have $\mu
= \lambda = \lambda_{\frac12}$ while $\bmu = \lambda_0$ is trivial. Hence, the
head of the operators must be taken from sections in the bundle
\begin{equation} \label{Gamma12}
\Gamma_{\lambda_{\frac12}} = \Lambda_{0,1} +
\sum_{l=1}^\infty \Lambda_{l,0} + \mbox{typicals}\ .
\end{equation}
The decomposition on the right hand side is obtained from the
formulas in appendix \ref{app:osp42_osp32}, just as in the previous example.
We see that one $\frac12$BPS section in the bundle of the fundamental
representation $\lambda_{\frac12}$ of $\ospH$ is the adjoint multiplet
of $\ospG$. The corresponding fields are the Noether currents. According
to our result \eqref{eq:oneloopdimension} their one-loop anomalous dimension
vanishes since both the Casimir of the fundamental $\lambda_{\frac12}$
and the Casimir of the adjoint $\Lambda_{0,1}$ vanish. The remaining
$\frac12$BPS fields are derivatives of the zero modes. Their anomalous
dimension is the same as for the zero modes themselves.

The $\frac12$BPS spectrum of operators of weight $(h_\infty,\bar h_\infty)=(1,1)$ is
a bit richer. In this case, our operators must contain $j$ and $\overline{\jmath}$
so that $\mu = \lambda_{\frac12} = \bmu$. In the tensor product of the two
fundamental representations $\mu$ and $\bmu$ we find $\lambda =\lambda_0$, $[1,0]$,
$[\frac12,1]$. Hence, the zero mode contributions can come from 3 different bundles.
The decomposition of the bundle $\Gamma_0$ was described in eq.\ \eqref{Gamma0}
already. So it remains to describe the two bundles
\begin{equation}
\Gamma_{[1,0]} = 2\Lambda_{0,1} + \Lambda_{0,2} + \mbox{typicals}
\end{equation}
and
\begin{equation}
\Gamma_{[\frac12,1]} =   \sum_{l=2}^\infty
(2 \Lambda_{l,0} + \Lambda_{l,1} + \Lambda_{l,-1}) + \mbox{typicals}\ .
\end{equation}
If we sum up all the contributions from the three possible bundles, we
find that the spectrum of operators of weight $(h_\infty,\bar h_\infty)
= (1,1)$ decomposes into
\begin{equation}\label{11operators}
\Gamma_{\lambda_\frac12 \otimes \lambda_\frac12} \cong
\Lambda_{0,0} + 2\Lambda_{0,1} + \Lambda_{0,2} + \Lambda_{1,0} +
\sum_{l=2}^\infty (3 \Lambda_{l,0} + \Lambda_{l,1}+\Lambda_{l,-1})
+ \mbox{typicals}
\end{equation}
The one-loop anomalous dimension of the corresponding operators is
determined by the first label of the representation,
\begin{equation}
\delta^{(1)}_R h = \frac{1}{2R^2} \Cas_{\g}(\Lambda_{l,k}) =
\frac{l^2}{2R^2}\ .
\end{equation}
We see in particular that our sigma model contains $145$ operators
with vanishing one-loop anomalous dimension. These sit in four different
representations of $\ospG$. There is one state in the trivial
representation $\Lambda_{0,0}$. This is the sigma model interaction
that remains marginal at one-loop. It actually remains marginal at all
loops. In addition, there are two adjoint multiplets $\Lambda_{0,1}$
of dimension $17$ each. The multiplicity two is actually a signature
of the distinction between projective covers and irreducibles. As we
explained above, one could have expected that the multiplicity of the
adjoint $\ospG$ section in the bundle associated to the adjoint
representation $[1,0]$ of $\ospH$ is given by the number of times
the 12-dimensional $[1,0]$ appears in the decomposition of the
17-dimensional $\Lambda_{0,1}$. Clearly, this multiplicity is one
which is not the correct answer for the number of $\Lambda_{0,1}$
multiplets in $\Gamma_{[1,0]}$. So indeed the example illustrates
nicely how important it is to determine the multiplicity of short
operators using decompositions of projective covers rather than
irreducibles.

\subsection{An alternative construction of vertex operators}
\label{sec:alt_constr_of_vertex_ops}

In order to fully appreciate the results of the previous subsection
and the elegance of their derivation, we would like to compare our
findings with more conventional constructions of vertex operators
from the fundamental field multiplet $X$. In doing so, we will have to
struggle a little bit with the implications of the constraint
\eqref{constraint} and the equations of motion \eqref{eom} on
counting coset fields. As a reward, we will understand e.g. that
the number $145$ of operators with vanishing one-loop anomalous
dimension contains non-trivial information about the dynamics of the
supersphere sigma model.

In building coset  fields from the fundamental field multiplet $X$
we shall start with the zero modes. For $h_\infty=\bar h_\infty = 0$
the relevant fields contain no derivatives and they are given by
monomials $F_{l,0}(X)$ of order $l = 0,1,2, \dots$ in the components
of $X$. Once we implement the constraint $X^2=1$ the components of
$F_{l,0}(X)$ transform in the traceless symmetric tensor
representations $\Lambda_{l,0}$. This agrees with our formula
\eqref{Gamma0} above.

Let us now proceed to fields of weight $(h_\infty, \bar h_\infty) =(1,0)$.
These must be of the form
\begin{equation}  \label{field10}
F_{l,0}(X) \,\partial X
\end{equation}
for $l=0,1,2,\dots$. The space of such objects transforms in the
tensor product $\Gamma_0 \otimes \Lambda_{1,0}$ of symmetric traceless
tensors with the fundamental $\Lambda_{1,0}$. But not all these fields
are non-zero. In fact, by taking a derivative of the constraint
$X^2 =1$ we obtain
\begin{equation}
X \cdot \partial X = X_a \partial X^a = 0
\end{equation}
Consequently any field of the form $F_{l,0} X \cdot \partial X$
vanishes. Such fields transform in the representation $\Gamma_0$.
If we remove them from the list \eqref{field10} we end up with a
space of fields transforming in
\begin{equation}\Gamma_{0} \otimes \Lambda_{1,0} - \Gamma_0  = \Lambda_{0,1} +
\sum_{l=1}^\infty \Lambda_{l,0} + \text{typicals} = \Gamma_{\lambda_\frac12}\ .  
\end{equation}
This agrees with our result \eqref{Gamma12}. We have already
interpreted the corresponding fields as the Noether currents and
derivatives of the zero modes.

Let us now turn to the most interesting set of fields, those with
weights $h =1 = \bar h$. In this case, the counting will be
affected by the equations of motion. The relevant fields
can all be written in either of the following forms
\begin{equation} \label{families}
 F_{l,0}(X)\, \partial \bar \partial X \quad , \quad
F_{l,0}(X) \, \partial X \bar \partial X\ .
\end{equation}
Our analysis of the space of these operators will proceed in two
steps. First we shall fully implement the constraint $X^2 =1$ and
then we consider the equations of motion. By taking derivatives of
the constraint $X^2=1$ we obtain the two equations
\begin{equation} X \cdot  \partial X  = 0 = X \cdot
\bar \partial X \  . 
\end{equation}
We can multiply each of these two equations with one of the
previously found operators of dimension $(h_\infty,\bar  h_\infty)=(1,0)$
or $(h_\infty,\bar h_\infty) =(0,1)$, respectively. All such operators
vanish. As we discussed above, they transform in $2\Gamma_{\lambda_\frac12}$.
Additionally, we also need to remove all operators created
from the zero modes by multiplication with the operator
$X\cdot \partial X X \cdot \bar \partial X$. These transform in
$\Gamma_{0}$.
This is not quite the end of story.
In fact, there is another family of operators that vanishes because
of the constraint $X^2 =1$. To see this, we differentiate the
constraint $X^2 =1$ by  $\partial \bar\partial$ and obtain
\begin{equation} \partial X \cdot \bar\partial X =
     -X \cdot \partial \bar\partial X \ .  
\end{equation}
This constraint allows us to remove all the operators of the
form $F_{l,0} \partial X \cdot \bar\partial X$. In other words
when considering the second family in eq.\ \eqref{families}, we
can restrict to those operators for which $\partial X \bar \partial X$
transforms wither in the representation $\Lambda_{2,0}$ (symmetric
traceless) or in $\Lambda_{1,0}$ (antisymmetric). Putting all this
together we find
\begin{eqnarray*}
	& & \hspace*{-2cm} \Gamma_0 \otimes \Lambda_{1,0} + \Gamma_0 \otimes (\Lambda_{2,0}
	+ \Lambda_{0,1}) - 2\Gamma_{\lambda_\frac12}-\Gamma_{0}\\[2mm]  &=&
	\Lambda_{0} + 3 \Lambda_{0,1} + \Lambda_{0,2} + 2\Lambda_{1,0}
+ \sum_{l=2}^\infty (4\Lambda_{l,0} + \Lambda_{l,1}+ \Lambda_{l,-1})\\[0mm]
& = &\Gamma_{\lambda_\frac12 \otimes \lambda_\frac12} +
\Lambda_{0,1} + \sum_{l=1}^\infty \Lambda_{l,0}.
\end{eqnarray*}
A quick glance at eq.\ \eqref{11operators} shows that we obtained more
than we expected. The reason is simple. While we have correctly implemented
the constraint $X^2 =1$, operators of weight $(h_\infty,\bar h_\infty) =(1,1)$
are the first ones to be sensitive to the equations of motion. The latter
precisely remove the unwanted multiplets. In the block of the zero, for example,
the operators
\begin{equation} X_I \partial \bar\partial X_J  - X_J \partial \bar \partial
X_I  
\end{equation}
contribute one of the three $\Lambda_{0,1}$ in the decomposition
we have listed. Once we insert the equations of motion, however,
these operators are set to zero
\begin{equation} X_I \partial \bar\partial X_J  - X_J \partial \bar \partial
X_I = \partial X\cdot \bar\partial X  \ (X_I X_J - X_J X_I)= 0 \ . 
\end{equation}
Hence, the fact that we found 145 operators of weight $(h_\infty,
\bar h_\infty) = (1,1)$ with vanishing one-loop anomalous dimension is
sensitive to the  equations of motion. Without them there would be 17
additional such operators.

\section{Duality with \texorpdfstring{$\ospft$}{osp(4|2)} Gross-Neveu Model}
\label{sec:duality_with_Gross-Neveu}

One lesson which has been learned through past studies of sigma models
is that they should not be considered as an isolated research topic.
There exist other important constructions of 2D (conformal) field
theories which are intimately tied to sigma models and sometimes can
provide intriguing insights into the non-perturbative features of sigma
models. We have already alluded to the example of sigma models on
Calabi-Yau spaces which possess a dual description in terms of (products
of) WZNW coset models. Another, more elementary, example is the compactified
free boson which admits a dual description in terms of two Majorana fermions.
The proposed duality between the sigma model on $S^{3|2}$ and the $\ospft$
Gross-Neveu model that we described in the introduction is quite similar
to the Coleman-Mandelstam duality between bosons and fermions only that
the abelian symmetry $\mathfrak{u}(1) = \mathfrak{so}(2)$ is replaced by
the non-abelian $\ospft$.

In the first subsection we shall describe the $\ospft$ Gross-Neveu model
and some of its most basic features. Then we review a central all-loop
result from \cite{Candu:2012xc} on the (target space) $\frac12$BPS spectrum
of perturbed supergroup WZNW models and explain how it applies to the
$\ospft$ Gross-Neveu model. In the third subsection we try to match the
$\frac12$BPS spectrum of the Gross-Neveu model for a certain value of
the Gross-Neveu coupling to the one-loop spectrum of the supersphere sigma
models. We will find perfect agreement for low lying states, but also
some discrepancies that involve fields with more derivatives. The
discussion of these findings is mostly deferred to the final section.

\subsection{The \texorpdfstring{$\ospft$}{osp(4|2)} Gross-Neveu model}

The fundamental field multiplet $\Psi = (\Psi_A) = (\psi_j,\gamma_a)$ of the $\ospft$
Gross-Neveu model consists of four Majorana fermions $\psi_j,j=1,\dots, 4,$ and a
bosonic $\beta\gamma$-system whose fields we shall denote by  $\gamma_1 = \gamma$
and $\gamma_2 = \beta$. In addition, there is a second multiplet $\overline \Psi = (\overline\psi_j,\overline \gamma_a)$ of opposite chirality. All these six fields
in $\Psi$ possess conformal weight $h=1/2$ and transform in the fundamental
representations $\Lambda_{1,0}$ of $\ospft$. The same applies to $\overline\Psi$.
In terms of these field multiplets, the action of the Gross-Neveu model reads
\begin{equation}\label{GNact}
	\begin{split}
		{\mathcal S}^{\text{GN}}[\psi,\gamma,\overline\psi,\overline\gamma] &=
		\frac{1}{2\pi} \int d^2z \biggl[
			{\sum}_j \bigl(\psi_j \bar\partial \psi_j +
			\bar{\psi}_j \partial \bar{\psi}_j\bigr)
			+ \bigl(\gamma_2 \bar \partial \gamma_1 + \bar{\gamma}_2 \partial
			\bar{\gamma}_1\bigr) \biggr] \\[2mm]
		 & \hspace*{2cm} + \frac{g^2}{2\pi} \int d^2z \biggl[{\sum}_j
				\psi_j \bar{\psi}_j + (\gamma_1
				\bar{\gamma}_2 - \gamma_2 \bar{\gamma}_1)\biggr]^2\ .
			\end{split}
\end{equation}
The $\ospft$ invariance of this action is manifest since all indices are
contracted with the $\ospft$ invariant metric. When written in terms of
$\Psi$ and $\overline\Psi$, rather than its components, the action takes
the same form as that of the massless Thirring model with its characteristic
fourth order interaction term. When the coupling constant $g$ is set to zero
the model is free and scale invariant. It possesses a Virasoro symmetry with
central charge $c=1$. The latter receives a contribution $c_j = 1/2$ from
each of the fermions $\psi_j$ and $c_a =-1/2$ from the two components of the
$\beta\gamma$-system. Switching on the coupling $g$ introduces a very
non-trivial action but it turns out to preserve conformal symmetry. In fact,
the $\beta$-function for the coupling $g$ is proportional to the dual
Coxeter number  $h^\vee = \Cas_{\g}(\Lambda_{0,1})$ and hence vanishes for
$\ospft$. Therefore, the $\ospft$  Gross-Neveu model defines a one-parameter
family of interacting conformal field theories with central charge $c=1$.

While the interaction in the $\ospft$ Gross-Neveu model preserves the Virasoro
and a global $\ospft$ symmetry, the free field theory possesses additional
current algebra symmetries that are broken when $g \neq 0$. In order to
describe these symmetries, we recall that the components of the field
multiplet $\Psi$ obey the following operator product expansions
\begin{equation} \psi_i(z) \psi_j(w) \sim \frac{\delta_{ij}}{z-w} + \dots \quad ,
\quad \gamma_{2}(z) \gamma_{1}(w) \sim \frac{\delta_{ab}}{z-w}\ .  
\end{equation}
Using these operator product expansions between the fundamental constituents
it is standard to show that the following quadratic combinations
\begin{equation} \label{currents}
J_{AB} = \Psi_A \Psi_B \quad \mbox{ where } \quad
(\Psi_A) = (\psi_i,\gamma_b)
\end{equation}
obey the algebraic relations of an $\ospft$ current algebra at level
$k=1$. Let us stress once again that this current algebra symmetry is
broken as soon as we switch on the coupling.

The current algebra symmetry suggests interpreting the free theory at $g=0$
as a Wess-Zumino-Novikov-Witten (WZNW) model. In addition, it is not difficult
to verify that the fourth order interaction term of the Gross-Neveu model can
be expressed in terms of the currents \eqref{currents} as
\begin{equation} \label{int}
 \frac{g^2}{2\pi} \int d^2z \left[{\sum}_i
 \psi_i \bar{\psi}_i + \gamma_1
\gamma_2 - \gamma_2 \bar{\gamma}_1\right]^2
\ = \ \frac{g^2}{2\pi} \int d^2z \sum_{AB} J_{AB}(z) \bar
J^{AB}(\bar z) \ .
\end{equation}
Putting all this together we have shown that the Gross-Neveu model can
be thought of as a deformed WZNW model at level $k=1$,
\begin{equation}
{\mathcal S}^{\text{GN}} = {\mathcal S}^{\text{WZNW}}_{k=1}  +
\frac{g^2}{2\pi} \int d^2z \sum_{AB} J_{AB}(z) \bar
J^{AB}(\bar z) \
\end{equation}
with the deformation being generated by an exactly marginal current-current
interaction. This reformulation of the $\ospft$ Gross-Neveu model will become
important when we apply the powerful results of \cite{Candu:2012xc} to the
Gross-Neveu model.

\subsection{An all-loop result for deformed WZNW models}

In \cite{Candu:2012xc}, current-current deformations of supergroup WZNW
models were studied. In particular it was argued that the deformation
by the operator
\begin{equation}
	\Omega(z,\bar{z})=J^{\mu}(z)\bar{J}_{\mu}(\bar{z})\ .
	\label{eqn:WZW_deformation}
\end{equation}
is truly marginal, provided that the Lie supergroup possesses vanishing
dual Coxeter number, i.e.\ that $G = \text{PSL}(N|N)$, $\text{OSP}(2N+2|2N)$,
$\text{D}(2,1;\alpha)$. In the definition of $\Omega$ the sum runs over all
directions $\mu$ in the Lie superalgebra $\mathfrak{g}$. The deformation
breaks the affine symmetry. Since it does not even commute with the zero
modes of the chiral currents, it also breaks the left and right
$\mathfrak{g}$ symmetries.
On the other hand, the sum  of left and right zero
modes does commute with the perturbing operator so that the deformed
theory preserves the diagonal $\mathfrak{g}$ action.

Of course, under the perturbation with the operator \eqref{eqn:WZW_deformation}
the conformal weight of fields can change, i.e.\ fields may develop an
anomalous dimension. In general, this anomalous dimension is difficult
to compute, at least beyond the leading order in perturbation
theory. Remarkably, for a special subset of fields, the authors of
\cite{Candu:2012xc} managed to obtain an all order expression. In physics
terminology, the fields for which this was possible are those that
transform in maximally atypical, or $\frac12$BPS, representations of the target
space symmetry $\g$. More precisely, the formulas of \cite{Candu:2012xc}
hold for all indecomposable field multiplets of $\mathfrak{g}$ which
contain a subrepresentation of non-zero superdimension. For such fields,
the anomalous dimension reads
\begin{equation}
	\delta^{(\infty)}_g h_{\text{BPS}} = \frac{g}{2(1-k^2g^2)}
\left[\Cas^D_{\g}(\Lambda_\text{BPS}) - (1-kg)\Bigl(\Cas_{\g}^L +
	\Cas_{\g}^R\Bigr)\right]\ .
	\label{eqn:anom_dim_gen_formula}
\end{equation}
Here $\Cas^{L/R}_{\g}$ refers to the value of the quadratic Casimirs on
the left and right representations in the unperturbed model, respectively.
The superscript $D$ means that the Casimir element is evaluated with respect to
the
diagonal action. We have placed the subscript 'BPS' on both sides of the
equation to remind us that this formula should only be applied to fields
that transform in maximally atypical representations $\Lambda$ under the
diagonal action. On the other hand, their transformation law with respect
to left or right action in the WZNW model is not constrained.

Let us now specialize this very general result to the $\ospft$ Gross-Neveu
model or, equivalently, to the current-current deformation of the $\ospft$
WZNW model at level $k=1$. In this case our formula can be applied to all
fields that transform in one of the atypical representations $\Lambda_{l,k}$
or any indecomposable composites formed from these. Let us recall that
the value of the quadratic Casimir element assumes the value $\Cas_{\g}
(\Lambda_{l,k}) = l^2$ on such atypicals. Hence, our general formula
\eqref{eqn:anom_dim_gen_formula} becomes
\begin{equation}
	\delta^{(\infty)}_g h_{\text{BPS}} = \frac{gl^{2}}{2\left( 1-g^{2} \right)} -
	\frac{g}{2\left( 1+g \right)}\left( \Cassq^{L} + \Cassq^{R} \right).
	\label{eqn:anom_dim_gn_formula}
\end{equation}
for fields transforming in $\Lambda = \Lambda_{l,k}$ with respect to the
diagonal action of $\g$. Note that the function $\delta^{(\infty)}_g h$
develops a singularity at $g=-1$, at least for a large number of states.
This simple observation motivates the
identification of  the point $g=-1$ with the $R\to\infty$ limit of the
$S^{3|2}$ sigma model. In fact, in the sigma model one expects that all
winding states develop infinite energy when $R \to \infty$. So, if we
want the sigma model to be dual to the Gross-Neveu model, we are forced
to identify $g=-1$ with the infinite radius limit. The precise relation
between the coupling $g$ and the radius $R$ reads \cite{Candu:2012xc}
\footnote{The cohomological methods developed in \cite{Candu:2010yg}
imply that the relation is identical to the one that appears in the
duality between a compactified free boson and the massless Thirring
model.}
\begin{equation}
	g=\frac{4-R^{2}}{4+R^{2}}.
	\label{eqn:g_R_ident}
\end{equation}
For a state to remain in the spectrum at the point $g=-1$, the
anomalous dimension \eqref{eqn:anom_dim_gn_formula} has to remain
finite. This is the case if
\begin{equation}
	\Cassq^{L}+\Cassq^{R} = \frac{l^{2}}{2}.
	\label{eqn:no_winding}
\end{equation}
We call eq. \eqref{eqn:no_winding} the no-winding condition. For states
that satisfy this condition, the anomalous dimension \eqref{eqn:anom_dim_gn_formula}
simplifies to
\begin{equation}
	\delta^{(\infty)}_g h_{\text{BPS}}  =
    \frac{1}{4}\frac{gl^{2}}{1-g}=
	-\frac{l^2}{8}+\frac{l^2}{2R^{2}} \ .
	\label{eqn:anom_dim_no_winding}
\end{equation}
Here we also inserted eq.\ \eqref{eqn:g_R_ident} so that the anomalous
dimension of the Gross-Neveu model fields is finally written in terms of
the radius parameter $R$ on the sigma model.
We have now gathered all the ingredients we need in order to perform
our first tests of the duality. Eq.\ \eqref{eqn:no_winding} tells us
which states of the free field theory make it into the spectrum at
$g=-1$ and eq.\ \eqref{eqn:anom_dim_no_winding} allows us to
compute the corresponding conformal weight. We will now start to
compare the resulting spectrum at $g=-1$ with the free supersphere
sigma model.

In our discussion of the one-loop anomalous dimensions for coset sigma
models we briefly commented on a puzzling instability that arises from
high gradient operators. The same type of instabilities also appears in
perturbed WZNW models, at least for generic choices of the target group
and the level. To leading order in perturbation theory this was observed
by Ryu et al. in \cite{Ryu:2010iq}. With the help of formula
\eqref{eqn:anom_dim_gen_formula} one may show that these instabilities
persist to any order in perturbation theory. The authors of \cite{Ryu:2010iq}
also observed that no instabilities occur for $\mathfrak{psu}(N|N)$
WZNW models at level $k=1$. This observation, however, does not carry
over to our $\ospft$ WZNW model at level $k=1$. In fact, one can show
that this theory contains instabilities arbitrarily close to the free
field theory, much as it is the case for sigma models. For now, we
shall close an eye on these issues.

\subsection{Checking the proposed duality}

We want to apply the results on the deformation of the $\frac12$BPS spectrum
in deformed supergroup WZNW models in order to test the proposed duality between
the $\ospG$ Gross-Neveu model and the supersphere sigma model. In the first
subsection we shall show that the zero mode spectrum of the sigma model is
recovered along with its 1-loop deformation. This is a remarkable example of
an emergent geometry. In the WZNW model, the fields that are associated with
spherical harmonics of the supersphere possess very large scaling dimensions.
These come down until they become zero modes, i.e.\ fields with vanishing
scaling weight, in the sigma model limit. Let us anticipate that the singular
vectors of the $\ospG$ WZNW model at level $k=1$ play an important role for
this identification with the zero mode spectrum of the sigma model to work
out. Then we turn to derivative fields of the sigma model. We will argue that
the agreement continues to hold for fields of conformal weight $(h_\infty,\bar h_\infty)
= (1,0), (0,1)$ in the sigma model. This may not come as a big surprise. Things
become more interesting for the fields with conformal weight $(h_\infty,\bar h_\infty)=
(1,1)$ since these are sensitive to the equations of motion in the sigma model.
Recall that in the sigma model we found $145$ states with vanishing 1-loop
scaling dimension. This will be exactly matched by the deformed WZNW model.
In the WZNW model, the scaling dimension of the corresponding $145$ states
is independent of the coupling so that the conjectured duality makes an
interesting prediction: All higher loop corrections to the scaling weight
of the $145$ states are actually zero. The match between the deformed WZNW
model and the sigma model extends to many other fields with $(h_\infty,\bar
h_\infty) = (1,1)$. On the other hand, we will also find sigma model fields
that cannot be reproduced within the deformed WZNW model.

\subsubsection{Ground state spectrum}

One key piece of evidence in support of the proposed duality
is the observation that we can actually recover all the zero
modes of the sigma model. Under the action of the global
\ospft\ symmetry the space $\Gamma_0$ of functions on the
supersphere decomposes into a sum of irreducible multiplets
$\Lambda_{l,0}$, see eq.\ \eqref{Gamma0}. Each of these multiplets
appears with multiplicity one. Other atypical representations
	$\Lambda_{l,k}$, $k\neq0$ do not occur.

As we have explained before, the states of the Gross-Neveu model
are constructed from a chiral multiplet $\Psi = \Psi^L$ that
transforms in a 6-dimensional representation of $\ospG$. The
$\ospG$ representation matrices are those known from the usual
fundamental representation, but the grading rules are reversed
so that the fermionic subspace is 4-dimensional while the
bosonic has dimension 2. It is a remarkable fact that the
conformal dimension $h$ of all chiral operators $\mathcal{O}^L$
in the undeformed case is bounded from below by
\begin{equation}
	h_0\bigl(\mathcal{O}^L_{[\Lambda]}\bigr) \geq
\frac{1}{2}\Cas_{\g}^L(\Lambda) \ .
	\label{eqn:hmin_bound}
\end{equation}
for all $\mathcal{O}^L$ that transform in the representation $[\Lambda]$
with respect to the left $\ospft$ action. Of course, the corresponding
statement holds for all operators $\mathcal{O}^R$ that are constructed
from the components of $\overline{\Psi} = \Psi^R$ and their derivatives.
It is actually possible to establish the stronger lower bound
\begin{equation}
 h_0\bigl(\mathcal{O}^L_{[\Lambda]}\bigr) \geq j_1+j_2(j_2+1)+
 j_3(j_3+1)+|j_2-j_3|\geq \frac{1}{2}\Cas_{\g}^L(\Lambda)
 \label{eqn:hmin}
\end{equation}
which shows that the inequality \eqref{eqn:hmin_bound} can only be saturated
by very special multiplets, when $j_1=0,\frac12$. It turns out that for each
integer $l = 0,1,2,\dots$ there is a unique field multiplet $\mathcal{O}^L_l$ such
that
\begin{equation}
	h_0(\mathcal{O}^L_l) =  \frac{l^2}{2}.
	\label{eqn:hmin_bound_sat}
\end{equation}
The multiplet $\mathcal{O}^L_l$ is obtained as a graded symmetric
component in the $l$-fold tensor product of the fundamental. Since
our generating field multiplet $\Psi$ is fermionic, i.e. its grading
is reversed in comparison to the grading of the fundamental, the
multiplet $\mathcal{O}^L_l$ must contain $l(l-1)/2$ derivatives.
Hence, its conformal dimension $h(\mathcal{O}^L_l) = l/2 + l(l-1)/2
= l^2/2$.

Let us illustrate the construction of $\mathcal{O}_l^L$ with a few explicit
examples. Of course, the operator $\mathcal{O}^L_0$ is just the identity field
while $\mathcal{O}^L_1$ is the fundamental multiplet $\Psi$. The next
multiplet $\mathcal{O}^L_2$ appears at $h(\mathcal{O}^L_2) = 2$,
 \begin{equation} \mathcal{O}^L_2 = \left(\psi_A \partial \psi_B + (-1)^{|A||B|}
\psi_B \partial \psi_A\right) \ .  
\end{equation}
When we multiply the multiplet $\mathcal{O}^L_l$ with its anti-holomorphic
partner $\mathcal{O}^R_l$ we obtain a set of bulk fields which transform
in the product $\Lambda_{l,0} \otimes \Lambda_{l,0}$. The only component
that can satisfy the no-winding condition is the one in the representation
$\Lambda_{2l,0}$. Indeed,
\begin{equation}\Cas_{\g}(\Lambda_{2l,0}) = 4 l^2 = 2\bigl(\Cas^L_{\g}(\Lambda_{l,0})
+ \Cas^R_{\g}(\Lambda_{l,0})\bigr)\  .  
\end{equation}
Let us denote the this component of the product by $V_{2l} = V_{2l}(z,\bar z)$.
To summarize, we have now constructed a field multiplet $V_{2l}$ in the WZNW
model that transforms in the representation $\Lambda_{l,0}$ with respect to
both the left and the right action of $\ospft$ and in the representations
$\Lambda_{2l,0}$ with respect to the diagonal action. In the WZNW model, i.e.\
the free Gross-Neveu model, this field possesses weights $\bigl(h_0(V_{2l}),
\bar{h}_0(V_{2l})\bigr)= (l^2/2 ,l^2/2)$.

Since the field $V_{2l}$ transforms on the $\frac12$BPS representation
$\Lambda_{2l,0}$ of $\ospft$, we can apply the results of the previous
subsection to compute its dimension for any value of the coupling $g$
and in particular at the point $g=-1$. With the help of the leading
term in eq.\ \eqref{eqn:anom_dim_no_winding} we obtain
 \begin{equation} h(V_{2l})_{g=-1} = h_{0}(V_{2l}) - \frac{1}{8} 4 l^2 = 0. 
\end{equation}
Hence, we obtain precisely the spectrum provided by the spherical
harmonics $\Lambda_{2l,0}$ in the sigma model, i.e.\ at least one
half of the zero modes of the supersphere sigma model.
\footnote{One would expect to obtain the missing zero modes 
$V_{2l+1}$ from other sectors of the Gross-Neveu model. Without 
the inclusion of additional states, the Gross-Neveu model is 
related to an orbifold theory $S^{3|2}/\mathbb{Z}_2$ rather 
than the supersphere sigma model.} Remarkably,
this identification is also consistent with what we know about the
1-loop anomalous dimensions in the sigma model. In fact, if we
keep the next to leading term in eq.\ \eqref{eqn:anom_dim_no_winding}
we find
\begin{equation}
h(V_{2l})_g = \frac{l^2}{2} + \frac{g l^2}{1-g} =
\frac{2 l^2}{R^2} \ . 
\end{equation}
This should be compared with the result \eqref{1loopzeromodes} for the one
loop anomalous dimension of the sigma model vertex operators $V_{\Lambda_{2l,0},
\lambda_0}$. We see that also the 1-loop corrections to the scaling law agree.
In the deformed WZNW model, the formula \eqref{eqn:anom_dim_no_winding} is
actually exact, i.e.\ there are no further corrections by terms involving
higher powers of the sigma model coupling $1/R^2$. The duality therefore
predicts that the anomalous dimensions of zero mode fields in the sigma model
are 1-loop exact. It should not be too difficult to check this prediction
through a direct computation along the lines of \cite{Wegner:1987av,Wegner:1987gu},
where anomalous dimensions of tachyonic vertex operators in bosonic O(N) models
were computed up to four loops. The general structure of Wegner's results
suggest that higher order corrections indeed vanish for the conformal
supersphere models, but we have not yet completed an honest derivation.

Since our fields ${\cal O}^{L/R}_l$ are the only ones satisfying the bound
\eqref{eqn:hmin_bound} and the bulk field $V_{2l}$ the only fields we could
build from them that solve the no-winding condition \eqref{eqn:no_winding},
the deformed WZNW model contains no further field of weight $(h_\infty,
\bar h_\infty) = (0,0)$ at $g=-1$. Moreover, because of the bound
\eqref{eqn:hmin_bound}, all other WZNW fields that solve the no-winding
condition end up with $h_\infty + \bar h_\infty > 0$ for $g=-1$.
In the free sigma model, the conformal weights are determined by the number
of derivatives and hence they are certainly non-negative. So, our results
are in beautiful agreement with the proposed duality.

Let stress that the match of zero modes only works for the WZNW model
at $k=1$, i.e.\ it does make crucial use of the exact position of singular
vectors. In order to illustrate this point let us consider the space of
states ${\mathcal H}^{(l)}_k$ of conformal weight $h=2$ ($\bar h = 0$).
For an $\ospft$ WZNW model with $k > 1$, these transform in
\begin{equation} {\mathcal H}^{(2)} \cong  \Lambda_{0,1} + \Lambda_{0,1} \odot \Lambda_{0,1}
   = \Lambda_{0,0} +  \Lambda_{0,1} +
   \Lambda_{2,-1} + 2 \Lambda_{2,0} + \Lambda_{2,1}
		+ [2,0,0] \ .
\end{equation}
The term $\Lambda_{0,1}$ originates from the action of the modes $J^{AB}_{-2}$
while the term $\Lambda_{0,1} \odot \Lambda_{0,1}$  contains the contributions
of $J^{AB}_{-1} J^{CD}_{-1}|0\rangle$. A formula for the symmetric tensor product
$\odot$ of the adjoint $\Lambda_{0,1}$ can be found at the end of appendix
\ref{app:bg_osp42}.  Note that there appear four different multiplets in which the
Casimir element has the maximal value $\Cas_{\g} (\Lambda)= 4$, namely the multiplets
$\Lambda = \Lambda_{2,k}, k = 0,\pm 1$. At level $k=1$, the first singular vectors
appear at $h=2$ and these reduce the spectrum to
\begin{equation} {\mathcal H}^{(2)}_{k=1} \cong \Lambda_{0,0} +  \Lambda_{0,1}  +
    \Lambda_{2,0} + [2,0,0]\   
\end{equation}
so that the representations with maximal Casimir are reduced to a single one,
namely $\Lambda_{2,0}$. This is the unique multiplet in ${\mathcal H}^{(2)}_{k=1}$
that is used to build a zero mode at $g=-1$. WZNW models with level $k > 1$
contain many more zero modes and hence cannot be dual to the supersphere
sigma model.

\subsubsection{Spectrum of gradient operators}

After our success in matching the zero modes of the sigma model with
fields in the deformed WZNW theory, we want to move on to gradient fields
in the sigma model. Some of them are very easy to find. This applies in
particular to the operators of weight $(h_\infty,\bar h_\infty) = (1,0)$.
Their spectrum was described in eq.\ \eqref{Gamma12}. Most of these fields
emerge from the WZNW model derivative operators $\partial V_{2l}$ with $l=
1,2,\dots$. The fields $V_{2l}$ were constructed in the previous subsection.
The bulk operators $\partial V_{2l}$ have conformal weight $(h_0,\bar h_0)
= (l^2/2+1, l^2/2)$ and they transform in the representation $\Lambda_{2l,0}$.
By the same reasoning as above we obtain a family of fields with weight
$(h_\infty,\bar h_\infty) = (1,0)$ at the point $g=-1$ which transform
in the $\Lambda_{2l,0}$ representations of $\ospG$. Their 1-loop anomalous
dimension coincides with that of the corresponding zero modes. Of course,
the match with the operators of weight $(h_\infty,\bar h_\infty) = (1,0)$
is not surprising since they are obtained as derivatives in both the WZNW
and the sigma model description.

There is one more set of operators at $(h_\infty,\bar h_\infty)  = (1,0)$, namely
the Noether currents of the sigma model that sit in the representation $\Lambda_{0,1}$.
It is obvious that these arise from the  chiral currents $J^{AB}$ in the WZNW model.
In fact, the currents of the WZNW model transform in the representation $\Lambda^L
= \Lambda_{0,1}$ and $\Lambda^R = \Lambda_{0,0}$ with respect to the left and right
action of $\ospft$, respectively. Under the diagonal action, the transformation law
is described by the tensor product $\Lambda^D = \Lambda_{0,1} \otimes \Lambda_{0,0}
= \Lambda_{0,1}$. Since all these representations possess vanishing Casimir, the
no-winding condition \eqref{eqn:no_winding} is satisfied and the anomalous
contribution to the conformal weight vanishes. Hence, we can identify the
deformation of the WZNW currents with the Noether currents of the sigma model.

Let us now turn to the operators of conformal weight $(h_\infty,\bar h_\infty)=
(1,1)$ in the sigma model. Their spectrum in the sigma model is given by eq.\
\eqref{11operators}. Obviously, we can obtain some of these from the operators
$\partial \bar \partial V_{2l}, l=1,2,\dots$ in the WZNW model. But these fields
are not even close to exhausting content of eq.\ \eqref{11operators}. In
particular, the sigma model contains these $145$ marginal fields with
vanishing 1-loop anomalous dimension that we discussed extensively in
section \ref{sec:supersphere_spectrum} and so far we have not seen any of them.

These $145$ fields belong to mutiplets $\Lambda_{0,0} + 2 \Lambda_{0,1}
+ \Lambda_{0,2}$, all of which have vanishing Casimir. Hence, in the WZNW
model they must appear with $(h_0,\bar h_0) = (1,1)$. So, let us count the
fields in the WZNW model that have weights $(h_0,\bar h_0) = (1,1)$ and
vanishing Casimir. All of these fields must arise among $J_A \bar J_B$,
i.e.\ sit in the tensor product of the adjoint representation of $\ospG$
with itself. This tensor product is given by
\begin{equation} \Lambda_{0,1} \otimes \Lambda_{0,1} \cong
\Lambda_{0,0} +  2 \Lambda_{0,1} + \Lambda_{0,2}+
   \Lambda_{2,-1} + 2 \Lambda_{2,0} + \Lambda_{2,1}
		+ [2,0,0]\ .
\end{equation}
Indeed, this contains exactly $145$ fields in representations from the
block of the trivial representations for which the anomalous dimension
vanishes to all orders in the coupling and hence also around $g=-1$, in
perfect agreement with the sigma model results. Since the space of
marginal fields in the sigma model is truncated by the equations of
motion, the deformed WZNW model has the sigma model equations of
motion built in!

This is a remarkable agreement. On the other hand, looking back at
the sigma model spectrum \eqref{11operators} we realize that the
content of what looks like $\mathcal{P}_{\Lambda_{2l,0}}, l = 1,2,
\dots$ is still missing. Additional fields in these representations
that acquire weights $(h_\infty,\bar h_\infty)=(1,1)$ at $g=-1$
do exist in the WZNW, but these turn out not to match the 1-loop
data near $g=-1$. This is the first discrepancy between the
Gross-Neveu and the sigma model. We shall discuss this and
other discrepancies in more detail in the concluding section.

Before we do so, let us point out that,  once again, the singular
vectors are absolutely crucial in order for the WZNW model to
respect the sigma model equations of motion. As an example let
us look at the operators of the form $\partial\bar\partial V_{4}$.
These give rise to a single marginal sigma model field in the
representation $\Lambda_{4,0}$. If it was not for the singular
vectors of conformal weight $h=2$, the WZNW model would give many
more marginal fields in the same block. In fact, the tensor
product
 \begin{equation}(2\Lambda_{2,0} +  \Lambda_{2,1} + \Lambda_{2,-1})
\otimes
  (2\Lambda_{2,0} +  \Lambda_{2,1} + \Lambda_{2,-1})
\cong  \Lambda_{4,-2} + 4\Lambda_{4,-1} + 6\Lambda_{4,0} +
4 \Lambda_{4,1} + \Lambda_{4,2} + \dots
\end{equation}
where $+\dots$ stand for multiplets $\Lambda$ with
$\Cas_{\g}(\Lambda) < 16$, none of which satisfy the no-winding
condition. But those that do clearly outnumber the spectrum
of marginal sigma model fields.

\section{Conclusions}

In this work we have reviewed recent results on the spectrum of coset sigma
models and applied them to the conformal supersphere sigma model with target
space $S^{3|2}$. The example shows very clearly that the construction of
vertex operators designed in \cite{Candu:2013cga} provides easy access to
the spectrum of sigma models, at least to leading order in the sigma model
coupling. We have then used the results to test a conjectured dual
description of the sigma model on $S^{3|2}$ which becomes weakly coupled
deep in the strongly curved regime of the sigma model. The dual theory
may be regarded as an $\ospft$ Gross-Neveu model or, equivalently, a
deformed $\ospft$ WZNW model at level $k=1$. With the help of all-loop
results from \cite{Candu:2012xc} we were able to recover the zero mode
spectrum of the sigma model along with a number of gradient fields. In
particular, we argued that the sigma model equations of motion are
implemented in the deformed WZNW model.

There are quite a few open problems associated with both the perturbative
results we reviewed and with the duality. We have already explained the
issue of perturbative instabilities from high gradient operators in
sigma models, see the final remarks in section
\ref{sec:spectrum_of_coset_sm}. These remain puzzling and there
is a wide range of proposals on how they could be interpreted, including e.g.
the suggestion that they might be cured by non-perturbative effects
\cite{Castilla:1996qn}, or that they indicate the existence of higher
fixed points \cite{Polyakov:2005ss}. High gradient instabilities are not
limited to sigma models. In fact, they have also been observed to occur
in perturbed WZNW models \cite{Ryu:2010iq}. The authors of that work also
noticed that high gradient instabilities are avoided for deformed
$\mathfrak{psu}(N|N)$ WZNW models at level $k=1$ since in this case
singular vectors remove the unstable operators. This is not true for
$\mathfrak{osp}(2N+2|2N)$, however, which is plagued by high gradient
instabilities, even at level $k=1$. Since the phenomenon appears to
be so omnipresent, it seems mandatory to uncover its (ir)relevance.

The duality between the Gross-Neveu and the sigma model we studied
in section \ref{sec:duality_with_Gross-Neveu} also leaves us with a
number of interesting open questions. To begin with,
let us observe that for all states in the sigma model that are dual
to no-winding states of the WZNW model, the 1-loop anomalous dimension
must be exact, i.e. it  should not  receive any higher loop corrections.
We have actually stressed before that our formula \eqref{eqn:anom_dim_no_winding}
is exact, i.e.\ in its derivation we did not drop any terms of higher order in
$1/R^2$. The only $R$-dependent correction term agrees exactly with the 1-loop
result in eq.\ \eqref{eq:oneloopdimension}, assuming that $\Cas_{\h}(\mu) +
\Cas_{\h}(\bmu) =0$ and inserting $\Cas_{\g}(\Lambda_{l,k}) = l^2$. It would
be very interesting to verify
this consequence of the duality through a 2-loop computation. Some 2-loop
computations for high gradient operators in sigma models were
performed previously in \cite{Castilla:1996qn}. Of course, designing
an argument that establishes 1-loop exactness for the relevant
subsector in the sigma model would be even more remarkable.

In the last section we have also found some sigma model fields that do
not seem to possess a counterpart in the deformed WZNW model, namely a
large number of fields at weight $(h_\infty,\bar h_\infty)=(1,1)$. These
are not the only sigma model fields that cannot be matched. In fact, the
comparison of eqs.\ \eqref{eq:oneloopdimension} and
\eqref{eqn:anom_dim_no_winding} shows that fields for which the
sum $\Cas_{\h}(\mu)+\Cas_{\h}(\bmu) \neq 0$ cannot possess a counterpart
in the Gross-Neveu model, at least not in the sense we outlined. On the
other hand, there exist intriguing further coincidences between the spectra
of the two theories which we were not able to incorporate into the above
analysis. In particular, the authors of \cite{Mitev:2008yt} uncovered some
miraculous character identities that establish a correspondence between all
chiral fields in the sigma model, no matter how large $h_\infty$ or $\bar
h_\infty$, and fields in the deformed WZNW model. Unfortunately, the
one-loop data in the sigma model spoil this match. Of course, it is
possible that these discrepancies simply disprove the duality. On the
other hand, it seems somewhat tempting to speculate that the discrepancies
might have the same origin as the high gradient instabilities described
above. Very much in the spirit of \cite{Ryu:2010iq} one might hope that
the duality could even offer new insights into the instabilities, but so
far we have not been able to make this more concrete.

On a more technological level, our work demonstrates that existing results
on the spectrum of superspace sigma and WZNW models can provide very
powerful tools to test dualities and to develop an efficient description
of sigma models deep in the strongly coupled regime. There are many other
models to which these ideas might apply. In particular, a similar duality
between conformal sigma models on complex projective superspace and
$\mathfrak{psu}(N|N)$ WZNW models has been proposed at various places in the
literature, see \cite{CPdual}. It should also be possible to extend the
perturbative computations in superspace sigma models to those target
spaces that appear in the context of the AdS/CFT correspondence. This
requires two generalizations of the present setup. Whereas the 1-loop
results we have reviewed above are restricted to symmetric spaces $G/H$
in which $H \subset G$ is fixed by an automorphism of  order two, the
description of strings in AdS backgrounds involves subgroups $H \subset
G$ which are  held fixed by an automorphism of order four. The extension
to such generalized symmetric spaces is a bit cumbersome but should not
meet any fundamental difficulty. Another fundamental aspect of AdS
backgrounds is that they are non-compact. This has implications on the
way we construct normalizable sections, at least when the denominator
group $H$ is non-compact as well. For $AdS_2$ backgrounds, on the other
hand, the construction of vertex operators reviewed above remains
unaltered. We will address such compactifications in future research.

\section*{Acknowledgments}
The authors wish to thank Constantin Candu, Vladimir Mitev, Andreas Ludwig,
Christopher Mudry, Thomas Quella and Hubert Saleur for comments and
interesting discussions. The research leading to these results has received
funding from the People Programme (Marie Curie Actions) of the European
Union's Seventh Framework Programme FP7/2007-2013/ under REA Grant
Agreement No 317089 (GATIS).

\appendix

\section{Representation theory of \texorpdfstring{$\ospft$}{osp(4|2)}}
\label{app:bg_osp42}
In the following we give a very basic introduction to the Lie superalgebra
$\ospft$ and (some of) its finite dimensional representations. The complex
superalgebra $\g:=\ospft$ may be realized as the set of supermatrices,
\beqa\label{ospft}
\ospft=\left\{\left(\begin{array}{cc}A & B\\ J_{2}B^t &
D\end{array}\right): A^t=-A\text{ and }
D^tJ_{2}=-J_{2}D\right\} \ .
\eeqa
Here $A$ is a $4\times 4$ matrix, $D$ is a $2\times 2$ matrix and $B$ is rectangular
of size $4 \times 2$. In addition, we introduced the $2\times 2$ matrix $J_2=
\left(\begin{smallmatrix}0 & -1 \\ 1 & 0\end{smallmatrix}\right)$. As usual,
the Lie superalgebra $\g$ decomposes into an even, or bosonic, subalgebra $\g_{\bar{0}}=
\mathfrak{so}(4)\oplus\mathfrak{sp}(2)\cong\mathfrak{sl}(2)\oplus \mathfrak{sl}(2)
\oplus\mathfrak{sl}(2)$ and an odd, or fermionic, subspace $\g_{\bar{1}}$.

Our review of representations focuses on finite dimensional representations. As
usual for superalgebras, irreducible representations  fall into two different
categories. On the one hand, there are the generic long multiplets. These are also
known as typical representations in the more mathematical literature. On the other
hand, a superalgebra also possesses short or BPS multiplets which mathematicians
refer to as atypical representations. BPS multiplets can be put together into
indecomposable representations. We will only work with one class of such
indecomposables, namely the projective covers of atypical representations.

In order to make all this more precise, we note that an integral dominant
highest weight $\Lambda=(j_1,j_2,j_3)$ of $\g_{\bar{0}}$ is also one for
the full superalgebra $\g$ if it obeys the consistency conditions
\beqa
\label{consistency}
j_1=0\Rightarrow j_2=j_3=0\quad,\qquad j_1=\frac{1}{2}\Rightarrow j_2=j_3\ .
\eeqa
The ordering of our the spins $j_i \in \frac12\mathbb{Z}$ is such that the
the first spin is related to the symplectic subalgebra $\mathfrak{sp}(2)$
while the two others are associated with the orthogonal one. This is a bit
unfortunate but agrees with conventions in earlier literature. We shall use
the label $[\Lambda] = [j_1,j_2,j_3]$ to denote finite dimensional
irreducibles.

With these labels introduced we can now spell out the shortening conditions
we have mentioned above. A representation $[j_1,j_2,j_3]$ is atypical provided
the spins satisfy any one of the following conditions
\begin{equation}
  \begin{split}
\label{atypicality conditions}
2j_1&\ =\ -j_2-j_3\ ,\\[2mm]
2j_1&\ =\ j_2+j_3+2\ ,\\[2mm]
2j_1&\ =\ \pm(j_2-j_3)+1\ .
  \end{split}
\end{equation}
Otherwise the representation $[j_1,j_2,j_3]$ is typical. The eigenvalue of the
quadratic Casimir element in the irreducible representation $[\Lambda]$ is given
by
\begin{equation}
\label{Casimir primary} \Cas_{\g}(\Lambda)\ = \
-4j_1(j_1-1)+2j_2(j_2+1)+2j_3(j_3+1)\ \ .
\end{equation}
If the spins satisfy one of the shortening conditions \eqref{atypicality conditions}
the value of the quadratic Casimir element is a square, i.e. $\Cas_{\g}(\Lambda) = l^2$
with $l \in \mathbb{N}$. The atypical weights $\Lambda=(j_1,j_2,j_3)$, i.e. those
weights that satisfy one of the shortening conditions, can be divided into blocks
$\beta_l$ that contain all those representations $\Lambda \in \beta_l$ for which
$\Cas_{\g}(\Lambda)=l^2$. The corresponding atypical labels can be listed
explicitly \cite{Germoni2000:MR1840448},
\begin{equation} \label{atyp1}
  \begin{split}
\beta_0&\ =\ \left\{\Lambda_{0,0}=(0,0,0)\, ,\,
\Lambda_{0,k}=\frac{1}{2}(k+1,k-1,k-1)\, ,\,  k\geq 1\right\}\\[2mm]
\beta_l&\ =\ \left\{\Lambda_{l,k}\, ,\,  k\in \mathbb{Z}\right\}
  \end{split}
\end{equation}
where
\beqa   \label{atyp2}
\Lambda_{l,k}=\left\{\begin{array}{lr}\frac{1}{2}(-k+2, -k-l, -k+l)&
    \text{ if }k\leq -l\\[2mm]\frac{1}{2}(-k+1, k+l-1, -k+l-1) &\text{ if
    }-l+1\leq k\leq 0\\[2mm]\frac{1}{2}(k+1, k+l-1, -k+l-1)& \text{ if
    }0\leq k\leq l-1\\[2mm] \frac{1}{2}(k+2,k+l,k-l)&  \text{ if } l\leq k
   \end{array}\right. \ \ .
\eeqa
One sees easily, that the weights $\Lambda_{l,-k}$ for
$l\geq 1$ may be obtained from $\Lambda_{l,k}$ by simply
exchanging the second and the third Dynkin label. Furthermore, it
is possible to distinguish the weights $\Lambda_{l,k}$ according
to the atypicality condition \eqref{atypicality conditions} they
obey. The only weight to fulfill the first condition is
$\Lambda_{0,0}$. The weights belonging to the second condition are
$\Lambda_{0,k}$ for $k\geq 1$ and $\Lambda_{l,\pm k}$ for  $k\geq
l$. Finally, those the satisfy the last atypicality relation are
the $\Lambda_{l,\pm k}$ for $k< l$. In any case, each of the weights
fulfills at most one of the shortening conditions. This means that
all atypical representations of $\ospft$ possess the same degree of
atypicality, i.e. they are all what mathematicians refer to as
maximally atypical and physicists call $\frac12$BPS.

We can decompose all irreducible representations $[j_1,j_2,j_3]$ in terms
of irreducible subrepresentations of the bosonic subalgebra $\g_{\bar 0}$.
For typical representation one finds
\begin{equation}
  \begin{split}
\label{typical decomposition}
\left[j_1,j_2,j_3\right]_{g_{\bar{0}}}
  &\ \cong\ (j_1,j_2,j_3)\bigoplus_{\alpha,\beta=\pm\frac{1}{2}}
  (j_1-\frac{1}{2},j_2+\alpha,j_3+\beta)\\
  &\qquad\bigoplus_{\alpha=\pm 1}\big[(j_1-1,j_2+\alpha,j_3)
  \oplus(j_1-1,j_2,j_3+\alpha)\big]\oplus 2 (j_1-1,j_2,j_3)\\
&\qquad\oplus\bigoplus_{\alpha,\beta=\pm \frac{1}{2}}
(j_1-\frac{3}{2},j_2+\alpha,j_3+\beta)\oplus (j_1-2,j_2,j_3)\ \ .
\end{split}
\end{equation}
There are a few special cases for which the decomposition is not
generic. If $j_1\leq 2, j_2\leq 1$ or $j_3\leq 1$ then the above
decomposition formula must be truncated at the point where one or
more of the labels become negative. Moreover, there are two cases
for which the multiplicity of the $(j_1-1,j_2,j_3)$ submodule has
to be changed. If $j_1=1, j_2>0, j_3>0$ or $j_1>1,j_2=0,j_3>0$ or
$j_1>1,j_2>0,j_3=0$, then this block will appear only once and if
both $j_2$ and $j_3$ are null or $j_1=1$ and at least one between
$j_2$ and $j_3$ is null, then it will not be present at all. From
the decomposition into representations of the bosonic algebra we can
determine the dimension of typical representations
\beqa
\dim[j_1,j_2,j_3] = 16(2j_1-1)(2j_2+1)(2j_3+1) \ . \eeqa
The decomposition \eqref{typical decomposition} for $j_1\ge1$, is valid for
the indecomposable Kac modules that emerge when the spins $j_i$ satisfy one
of the shortening conditions \eqref{atypicality conditions}. These Kac
modules are composites of irreducibles. More precisely, one finds
\begin{equation}
  \begin{split}
\label{Kac modules}
&K_{\Lambda_{0,2}}:\ [\Lambda_{0,2}]\longrightarrow [\Lambda_{0,0}]\oplus [\Lambda_{0,1}]\\
&K_{\Lambda_{0,k}}:\ [\Lambda_{0,k}]\longrightarrow [\Lambda_{0,k-1}]\text{ for } k\geq 3 \\
&K_{\Lambda_{l,k}}:\ [\Lambda_{l,k}]\longrightarrow [\Lambda_{l,k-1}]\text{ for } k\geq 1\\
&K_{\Lambda_{l,k}}:\ [\Lambda_{l,k}]\longrightarrow
[\Lambda_{l,k+1}]\text{ for } k\leq -1\ \ .
  \end{split}
\end{equation}
The arrows mean that fermionic generators can take us from the representation
on the left to the one on the right but not vice versa. Put differently, the
representation on the right hand side of the arrows is a subrepresentation
of the Kac module. If we quotient the Kac module by this subrepresentation,
the corresponding factor representation is the one on the left hand side.
The representations with $j_1 = \frac12$ are somewhat special. In fact, when
$j_1 = \frac12$, the Kac module is irreducible and we obtain
 \beqa \left.{\Lambda_{l+1,2}}\right|_{\g_{\bar{0}}}=\left[\frac{1}{2}, \frac{l}{2},
\frac{l}{2}\right]_{\g_{\bar{0}}}\cong \left(\frac{1}{2},
\frac{l}{2}, \frac{l}{2}\right)\oplus \left(0,
  \frac{l+1}{2}, \frac{l+1}{2}\right)\oplus \left(0, \frac{l-1}{2},
  \frac{l-1}{2}\right)\ \ .
\eeqa
From our description of the Kac modules it is possible to determine the
dimensions of irreducible atypicals,
\begin{equation}
  \begin{split}
\label{atypical dimensions}
\dim [\Lambda_{0,0}]&\ =\ 1\ ,\qquad \dim [\Lambda_{0,1}]\ =\ 17\
,\qquad\dim [\Lambda_{l,0}]\ =\ 4l^2+2\\
\dim [\Lambda_{0,k}]&\ =\ (2k+1)\left[(2k+1)^2-3\right]\text{ for } k\geq 2\\
\dim [\Lambda_{l,k}]&\ =\ (2k+1)\left[4(l^2-1)-(2k+1)^2+7\right]\text{ for } k\leq l-1\\
\dim [\Lambda_{l,k}]&\ =\ (2k+3)\left[(2k+3)^2-4(l^2-1)-7\right]\text{ for } k\geq l\ .
\end{split}
\end{equation}
We are finally prepared to describe the projective covers that feature so prominently
in the construction of homogeneous vector bundles. While typical irreducibles $[\Lambda]$
coincide with their projective cover ${\mathcal P}_\Lambda= [\Lambda]$, the projective
cover of an atypical representations is an indecomposable composite of atypicals. Its
precise structure can be read off from the following diagrams
\beqa
&&\mathcal P_{\Lambda_{0,0}}:\Lambda_{0,0}\rightarrow\Lambda_{0,2}\rightarrow\Lambda_{0,0}
\label{projdec1}\\
&&\mathcal P_{\Lambda_{0,1}}:\Lambda_{0,1}\rightarrow\Lambda_{0,2}\rightarrow
\Lambda_{0,1}\label{projdec2}\\
&&\mathcal P_{\Lambda_{0,2}}:\Lambda_{0,2}\rightarrow\Lambda_{0,3}
\oplus\Lambda_{0,1}\oplus\Lambda_{0,0}\rightarrow\Lambda_{0,2}\label{projdec3}\\
&&\mathcal P_{\Lambda_{l,k}}:\Lambda_{l,k}\rightarrow\Lambda_{l,k+1}
\oplus\Lambda_{l,k-1}\rightarrow
\Lambda_{l,k}\qquad \text{otherwise} \label{projdec4}
\eeqa
The meaning of the arrows was explained in our discussion of Kac modules
above. Note that all the atypicals that appear in any given projective cover belong
to the same block $\beta$. It is actually not possible to build indecomposables
from representations within different blocks.

Before we conclude this brief overview over representations of the Lie superalgebra
$\ospft$ we want to spell out a few tensor product decompositions between irreducible
atypicals. These are used in our discussion of the low lying spectrum in the $\ospft$
Gross-Neveu model.
\begin{equation}
	\begin{aligned}
		\Lambda_{0,1}\otimes\Lambda_{0,1} &=
		\Lambda_{0,0}+ 2 \Lambda_{0,1}+ \Lambda_{0,2}+
		\Lambda_{2,-1}+ 2 \Lambda_{2,0}+ \Lambda_{2,1}
		+ [2,0,0]
		\\
		\Lambda_{0,1}\odot\Lambda_{0,1} &=
		\Lambda_{0,0}+ \Lambda_{2,-1}+ 2 \Lambda_{2,0}+ \Lambda_{2,1}
		+ [2,0,0]
		\\		
\Lambda_{0,1}\otimes\Lambda_{0,2} &=
		\Lambda_{0,0}+ \Lambda_{0,1}+ 3 \Lambda
		_{0,2}+ \Lambda_{0,3}
		+
		\\
		&+
		[1,1,1]+ [\tfrac{3}{2},\tfrac{1}{2},\tfrac{3}{2}]+
		[\tfrac{3}{2},\tfrac{3}{2},\tfrac{1}{2}]+ [2,0,1]+
		[2,1,0]+ [\tfrac{5}{2},\tfrac{1}{2},\tfrac{1}{2}]
		\\
		\Lambda_{0,2}\otimes\Lambda_{0,2} &=
		2 \Lambda
		_{0,0}+ 4 \Lambda_{0,1}+ 4 \Lambda
		_{0,2}+ 4 \Lambda_{0,3}+ \Lambda_{0,4}+
		\\
		&+\Lambda
		_{2,-2}+ 3 \Lambda_{2,-1}+ 4 \Lambda_{2,0}+ 3
		\Lambda_{2,1}+ \Lambda_{2,2}+
		\\
		&+\Lambda_{4,-1}+ 2
		\Lambda_{4,0}+ \Lambda_{4,1}
		+
		\\
		&+
		[1,0,2]+ 2
		[1,1,1]+ [1,2,0]+ 2
		[\tfrac{3}{2},\tfrac{1}{2},\tfrac{3}{2}]+ 2
		[\tfrac{3}{2},\tfrac{3}{2},\tfrac{1}{2}]+ 2
		[\tfrac{3}{2},\tfrac{3}{2},\tfrac{3}{2}]+
		\\
		&+ 2 [2,0,0]+ 2 [2,0,1]+ 2
		[2,1,0]+ [2,1,2]+ [2,2,1]+
		\\
		&+ 2 [\tfrac{5}{2},\tfrac{1}{2},\tfrac{1}{2}]+ 2
		[\tfrac{5}{2},\tfrac{1}{2},\tfrac{3}{2}]+ 2
		[\tfrac{5}{2},\tfrac{3}{2},\tfrac{1}{2}]+ [3,0,0]+
		[3,0,1]+ [3,1,0]+ [3,1,1]
		\\
		\Lambda_{1,0}\otimes\Lambda_{1,0} &=
		\Lambda_{0,0}+ \Lambda_{0,1}+ \Lambda_{2,0}
		\\
		\Lambda_{2,0}\otimes\Lambda_{2,0} &=
		\Lambda_{0,0}+ \Lambda_{0,1}+ \Lambda_{2,-1}
		+ 2 \Lambda_{2,0}+ \Lambda_{2,1}+
		\Lambda_{4,0}
		+  [1,1,1]
	\end{aligned}
	\label{eqn:app_tensorproducts}
\end{equation}
The + on the right hand side requires a short comment. As we have stated above, atypical
irreducibles can be combined to form larger indecomposables. This happens for many of the
atypical representations that appear in the above tensor product decompositions. Hence,
many of the atypicals are not direct summands. This is why we did not use $\oplus$. On
the other hand, the sum is direct for all projective modules, i.e.\ for typicals and
projective covers of atypicals. The symbol $\odot$ is used to denote the symmetric part
of the tensor product.

\section{Representation theory of \texorpdfstring{$\osptt$}{osp(3|2)}}
\label{app:osp32}

In this appendix we provide some background material on the Lie superalgebra $\osptt$
and its finite dimensional representations. The basic definition of $\osptt$ resembles
the definition \eqref{ospft} we gave for $\ospft$ only that now $A$ is a $3 \times 3$
matrix and $B$ is rectangular of size $3\times 2$. In the case of $\h= \osptt$, the
bosonic subalgebra is $\h_{\bar{0}}=\mathfrak{so}(3)\oplus\mathfrak{sp}(2)$. Since
$\h_{\bar{0}}$ has rank two, highest weights are labeled by two numbers $\lambda =
(q,p)$. In our conventions, the $\mathfrak{so}(3)$ spin $p$ runs over non-negative
integers while $q$ is a non-negative half-integer. Note that once again, the order
of the two labels is a bit unfortunate. As in the case of $\ospft$, there is an
additional constraint on the weights $(q,p)$ that must be satisfied in order for
$(q,p)$ to label a representation of $\osptt$, namely
$$ q=0\ \Rightarrow\ p=0\ . $$
Once more we shall use the bracket notation $[\lambda]= [q,p]$ to denote the associated
irreducible representation of $\osptt$. The representation $[q,p]$ is typical (long)
unless the labels $q,p$ satisfy one of the following two shortening conditions
\beqa
p+2q=0  \quad , \quad  p-2q+1=0 \ . \label{shorteningosptt}
\eeqa
These conditions are mutually exclusive. While the first one is only satisfied
for the trivial representation $q=p=0$, the latter singles out a one parameter
family of (maximally) atypical (or $\frac12$BPS) representations.

The eigenvalue of the quadratic Casimir element in an irreducible representation
$[\lambda]=[q,p]$ is given by
\beqa
\Cas_{\h}([q,p])\ = (p+2q)(p-2q+1)\ .
\eeqa
In particular, we conclude that the quadratic Casimir element vanishes for all
atypical representations of $\osptt$. This suggests that all atypicals belong to
one and the same block, which is indeed the case. Representations in this unique
block are given by
\beqa
\lambda_0=[0,0] \quad ,\quad  \lambda_q=[q,2q-1] \ .
\eeqa
Let us also mention in passing that the Lie superalgebra $\osptt$ possesses a
fourth order Casimir element whose eigenvalues are given by
\beqa
\label{Casimir4 osp32} \Cas^{(4)}_{\h}(\lambda)\ =\frac 14 \Cas_{\h}(\lambda)
[3p(3p+1)+2(q+1)(2q-3)]
\eeqa
The fourth order Casimir element does not show up in the 1-loop anomalous
dimensions but could enter starting from 2 loops.

As in the case of $\ospft$ it is useful to know how the irreducible representations
decompose with respect to the bosonic subalgebra. For typical representations, this
decomposition is given by
\beqa
[q,p]_{\h_{\bar 0}}\, \cong(q,p)\oplus
\bigoplus_{\alpha=0,\pm1}\left[(q-\tfrac12,p+\alpha)\oplus(q-1,p+\alpha)\right]
\oplus(q-\tfrac32,p)\ .
\eeqa
Truncations are present whenever one or both labels on the right hand side become
negative. When $q=\frac 12$ or $p=0$ the term $(q-\frac12,p)$ does not appear. For
the adjoint representation the decomposition reads
\begin{equation}
[1,0]_{\h_{\bar 0}} \cong (1,0)\oplus(\tfrac12,1)\oplus(0,1)\ .
\end{equation}
Note that in the case of $\osptt$ the adjoint representation is typical.
Atypical representations with $q\ge1$ possess the following decomposition
\beqa
[\lambda_q]_{\h_{\bar 0}}\,\cong(q,2q-1)\oplus(q-\tfrac12,2q-1)
\oplus(q-\tfrac12,2q)\oplus(q-1,2q)\ .
\eeqa
The atypical trivial representation $\lambda_0$ and the fundamental $\lambda_\frac12$
are special. While the decomposition of $\lambda_0$ is trivial, the fundamental
representation gives
\begin{equation}
[\lambda_{\frac12}]_{\h_{\bar 0}} \cong (\tfrac12,0)\oplus(0,1) \ .
\end{equation}
For completeness we also state the dimension of the these representations. In the
case of typical long multiplets we have
\begin{equation}
\dim\bigl([q,p]\bigr)=4(2p+1)(4p-1)
\end{equation}
while the dimension of atypicals is given by
\begin{equation}
	\begin{split}
		\dim[\lambda_0]=1\quad\dim[\lambda_{\frac12}]=5\\
		\dim[\lambda_q]=-2+32 q^2\ .
	\end{split}
\end{equation}
As for any Lie superalgebra, atypical representations can be combined into
larger indecomposables. For our analysis, the projective covers of atypicals
are of particular importance. Their structure is given by
\begin{align}
		\label{projdec21}
		\mathcal{P}_{\lambda_0}&:\lambda_0\rightarrow\lambda_1\rightarrow\lambda_0
		\\ \label{projdec22}
		 \mathcal{P}_{\lambda_{\frac12}}&:\lambda_{\frac12}\rightarrow\lambda_1\rightarrow\lambda_{\frac12}\\
		\label{projdec23}
		\mathcal{P}_{\lambda_1}&:\lambda_1\rightarrow\lambda_{\frac
		32}\oplus\lambda_{\frac12}\oplus\lambda_0\rightarrow\lambda_1\\
		\label{projdec24}
		\mathcal{P}_{\lambda_q}&:\lambda_q\rightarrow\lambda_{q+\tfrac12}
		\oplus\lambda_{q-\tfrac12}\rightarrow\lambda_q\qquad\text{otherwise}
\end{align}
The meaning of the arrows was explained in appendix \ref{app:bg_osp42}. The structure we
display here is consistent with the fact that all atypical irreducibles
$\lambda_q$ of $\osptt$ belong to the same block.

In our construction of coset vertex operators
\eqref{eq:fieldscosetmodel}, and in
particular in the analysis of the tail factors, we need some input about tensor
products of $\osptt$ representations. The first few powers of the fundamental
representation $\lambda_{\frac12}$ are given by
\begin{align}
\lambda_{\frac12}^{\otimes2}&= [1,0]+[\tfrac12,1]+\lambda_0
\\
\lambda_{\frac12}^{\odot2}&= [\tfrac12,1] +\lambda_0
\\
\lambda_{\frac{1}{2}}^{\odot3}&=[\tfrac12,2]+\lambda_{\frac12}
\end{align}
Here, we use the symbol $\odot$ to denote the graded symmetric part of the tensor
product. The formulas we displayed are relevant e.g.\ for products such as
$j \partial j$, $j^2$ and $j^3$, respectively. Let us also list a few
additional tensor products of low dimensional representations,
\begin{equation}
	\begin{split}
		[1,0]\otimes\lambda_{\frac12}&=
		[\tfrac32,0]+2\lambda_{\frac12}+\lambda_1
		\\
		\phantom{2}[\tfrac12,1]\otimes\lambda_{\frac12}&=
		[\tfrac12,2]+2\lambda_{\frac12} +\lambda_1
		\\
		\phantom{2}[\tfrac12,1]\otimes[\tfrac12,1]&=
		[1,2]+[1,0]+[\tfrac12,3]+[\tfrac12,1]+ 2\lambda_0+\lambda_1
		\\
		\phantom{2}[\tfrac12,2]\otimes\lambda_{\frac12}&=
		[1,2]+[\tfrac12,3]+[\tfrac12,1]
		\\
		\phantom{2}[\tfrac12,2]\otimes[\tfrac12,1]&=
		[1,3]+[1,2]+[\tfrac12,4]+[\tfrac12,2]+2\lambda_{\frac12}+\lambda_1
		\\
		\phantom{2}[\tfrac12,2]\otimes[\tfrac12,2]&=
		[1,4]+[1,3]+[1,2]+[1,0]+[\tfrac12,5]+[\tfrac12,3]
		\\
		&+[\tfrac12,2]+[\tfrac12,1]+2\lambda_0 +\lambda_1
		\\
		\phantom{2}[\tfrac12,1]\otimes[1,0]&=
		[\tfrac32,1]+[1,0]+[1,2]+[\tfrac12,1]
		\\
		\phantom{2}[\tfrac12,2]\otimes[1,0]&=
		[1,3]+[\tfrac12,2]+\lambda_{0}
		+\lambda_{\half}+2\lambda_{1}+\lambda_{\frac{3}{2}}
		\\
		\phantom{2}[1,0]\otimes[1,0]&=
		[2,0]+[\tfrac32,1]+[1,0]+[\tfrac12,1]+2\lambda_0+\lambda_1
	\end{split}
	\label{eqn:low_dim_osp32_products}
\end{equation}
These are useful in order to carry the construction of vertex operators
to higher gradient operators. Note that while it is not relevant for our
discussion, the atypical representations in
\eqref{eqn:low_dim_osp32_products} always combine into projectives,
while all other sums are direkt.

\section{Restriction of \texorpdfstring{$\ospft$}{osp(4|2)} representations to
\texorpdfstring{$\osptt$}{osp(3|2)}}
\label{app:osp42_osp32}

As we explained in section \ref{sec:homo_vec_bundles}, a key
ingredient in constructing vertex
operators on coset superspaces is the decomposition \eqref{Gammaexp} of
sections in homogeneous vector bundles into multiplets of the symmetry.
According to the central formula, the multiplicity $n_{\Lambda\lambda}$
of a $\mathfrak{g}$ multiplet $\Lambda$ in a bundle $\Gamma_\lambda$ is
given by eqn \eqref{fundlemma}. It implies that $n_{\Lambda\lambda}$ can
be computed through the decomposition
$$ \left.{\mathcal{P}}_\Lambda\right|_{\h} = \bigoplus_\lambda
    n_{\Lambda\lambda} \,{\mathcal{P}}_\lambda =
    \bigoplus_\lambda [   \left.{\mathcal{P}}_\Lambda\right|_{\h}\,:\,
    \mathcal{P}_\lambda] \ \mathcal{P}_\lambda \ .
$$
Given what we know about the projective covers of both $\ospft$ and $\osptt$
it is not too difficult to work out the multiplicities $n_{\Lambda\lambda}$.
We only need the results for atypical labels $\Lambda= \Lambda_{l,k}$. For
representations $\Lambda_{0,k}$ in the block of the trivial representation
one finds
\begin{equation}
	\begin{split}
		\mathcal{P}_{\Lambda_{0,0}}|_{\text{\tiny{$\osptt$}}}&=
		\mathcal{P}_{\lambda_0}\oplus[\tfrac32,0]\oplus[\tfrac{3}{2},1]
		\\
		\mathcal{P}_{\Lambda_{0,1}}|_{\text{\tiny{$\osptt$}}}&=
		\mathcal{P}_{\lambda_{\frac12}}\oplus[\tfrac32,0]
		\oplus[\tfrac{3}{2},1]
		\\
		\mathcal{P}_{\Lambda_{0,k}}|_{\text{\tiny{$\osptt$}}}&=
		\mathcal{P}_{\lambda_{\frac{k}2}}\oplus
		2\bigoplus_{n=0}^{k-1}[\tfrac{k+1}2,n]
		\oplus\bigoplus_{n=0}^{k}[\tfrac{k+2}2,n]
		\oplus\bigoplus_{n=0}^{k-2}[\tfrac{k}2,n]\ ,
		\quad \text{for all}\quad k\geq 2\ .
	\end{split}
	\label{pcdec1}
\end{equation}
Similarly one can decompose the projective covers of the
symmetric traceless tensor representations $\Lambda_{l,0}$,
\begin{equation}
	\begin{split}
		\mathcal{P}_{\Lambda_{1,0}}|_{\text{\tiny{$\osptt$}}}&=
		\mathcal{P}_{\lambda_0}\oplus\mathcal{P}_{\lambda_\frac12}
		\oplus2[\tfrac32,1]
		\\
		\mathcal{P}_{\Lambda_{2,0}}|_{\text{\tiny{$\osptt$}}}&=
		\mathcal{P}_{\lambda_0}\oplus\mathcal{P}_{\lambda_\frac12}
		\oplus2[\tfrac12,1]
		\\
		\mathcal{P}_{\Lambda_{l,0}}|_{\text{\tiny{$\osptt$}}}&=
		\mathcal{P}_{\lambda_0}\oplus\mathcal{P}_{\lambda_\frac12}
		\oplus2\bigoplus_{n=1}^{l-1}[\tfrac{1}2,n]\oplus2
		\bigoplus_{n=2}^{l-1}[1,n]\ , \quad \text{when}\quad l \geq 2\ .
	\end{split}
	\label{pcdec2}
\end{equation}
Finally, generic projective covers possess the following
decomposition into projectives of $\osptt$,
\begin{equation}
	\begin{split}
		\mathcal{P}_{\Lambda_{l,k}}|_{\text{\tiny{$\osptt$}}}=
		\mathcal{P}_{\lambda_{\frac {\abs{k}+1}2}}
		\oplus &\bigoplus_{n=\abs{k}}^{l-1}[\tfrac{\abs{k}}{2},n] \oplus
		2\bigoplus_{n=\abs{k}+1}^{l-1}[\tfrac{\abs{k}+1}{2},n] \oplus
		\bigoplus_{n=l}^{\abs{k}-1}[\tfrac{\abs{k}+1}{2},n]	
		\\
		\oplus\  2
		&\bigoplus_{\substack{n=l\\ \hphantom{n=\abs{k}}}}^{\abs{k}}
		[\tfrac{\abs{k}+2}{2},n]
		\oplus  \bigoplus_{n=\abs{k}+2}^{l-1}[\tfrac{\abs{k}+2}{2},n]\oplus
		\bigoplus_{n=l}^{\abs{k}+1}[\tfrac{\abs{k}+3}{2},n]
		\ .
	\end{split}
	\label{pcdec3}
\end{equation}
This last formula holds whenever $l \geq 1$ and $\abs{k} \geq 1$. Formulas
\eqref{pcdec1}-\eqref{pcdec3} provide the main input for the construction
of vertex operators in section \ref{sec:vertex_ops_and_anom_dim}.
Let us note that in these formulas all sums are direct since the
restriction of projective modules is a direct sum of projectives and
projectives cannot appear as pieces of larger indecomposibles.

In order to derive these decomposition
formulas one starts from the following decomposition formula for representations
of the bosonic subalgebra $\g_{\bar{0}}$ into representations of $\h_{\bar{0}}$,
\beq
(j_1,j_2,j_3)_{\h_{\bar{0}}} \cong \bigoplus_{p=\abs{j_2-j_3}}^{j_2+j_3}(j_1,p)
\eeq
In a second step these decomposition formulas are exploited to determine how
atypical irreducibles of $\ospft$ decompose upon restriction to $\osptt$. The
results read,
\begin{equation}
	\begin{split}
		\Lambda_{0,0}|_{\text{\tiny{$\osptt$}}}&=\lambda_0\\
		\Lambda_{0,k}|_{\text{\tiny{$\osptt$}}}&=\lambda_{\frac
		l2}\oplus\bigoplus_{n=0}^{k-1}[\tfrac{k+1}2,n],\,\,\, l>0\\
		\Lambda_{l,0}|_{\text{\tiny{$\osptt$}}}&=\bigoplus_{n=0}^{l-1}[\tfrac12,n]\oplus\lambda_0,\,\,\, l>0\\
		\Lambda_{l,k}|_{\text{\tiny{$\osptt$}}}&=\bigoplus_{n=\abs{k}}^{l-1}[\tfrac{\abs{k}+1}2,n], \,\,\,   0<\abs{k}\leq l-1\\
		 \Lambda_{l,k}|_{\text{\tiny{$\osptt$}}}&=\bigoplus_{n=l}^{\abs{k}}[\tfrac{\abs{k}}2+1,n]\oplus\lambda_{\frac{\abs{k}+1}2} ,\,\,\, 0 < l \leq \abs{k}
	\end{split}
\end{equation}
Since we know how projective covers are built from atypicals, it is now straightforward to
verify the decomposition formulas \eqref{pcdec1}-\eqref{pcdec3}.

\index{bulkspec1}


\end{document}